\documentclass[twocolumn]{aastex63}
\usepackage{amsmath}

\newcommand{\ba}{\begin{eqnarray}}
\newcommand{\ea}{\end{eqnarray}}
\submitjournal{ApJ}

\shorttitle{Unveiling IA with SC and DECaLS}
\shortauthors{Yao et al.}
\graphicspath{{./}{figures/}}
\begin{document}
	
	\title{Unveiling the Intrinsic Alignment of Galaxies with Self-Calibration and DECaLS DR3 data}

	\correspondingauthor{Ji Yao, Huanyuan Shan, Pengjie Zhang}
	\email{ji.yao@outlook.com; hyshan@shao.ac.cn; zhangpj@sjtu.edu.cn}
	
	\author[0000-0002-7336-2796]{Ji Yao}
	\affiliation{Department of Astronomy, Shanghai Jiao Tong University, Shanghai 200240, China}
	
	\author[0000-0001-8534-837X]{Huanyuan Shan}
	\affiliation{Shanghai Astronomical Observatory (SHAO), Nandan Road 80, Shanghai 200030, China}
	
	\author{Pengjie Zhang}
	\affiliation{Department of Astronomy, Shanghai Jiao Tong University, Shanghai 200240, China}
	\affiliation{Tsung-Dao Lee Institute, Shanghai Jiao Tong University, Shanghai 200240, China}
	\affiliation{Shanghai Key Laboratory for Particle Physics and
		Cosmology, China}
	
	%\author{Hu Zou}
	%\affiliation{National Astronomical Observatories, Chinese Academy of Sciences, 20A Datun Road, C%haoyang District, Beijing, China}
	
	\author{Jean-Paul Kneib}
	\affiliation{Institute of Physics, Laboratory of Astrophysics, Ecole Polytechnique Fédérale de Lausanne (EPFL), Observatoire
		de Sauverny, 1290 Versoix, Switzerland}
	\affiliation{Aix-Marseille Univ, CNRS, CNES, LAM, Marseille, France}
	
	\author[0000-0002-9253-053X]{Eric Jullo}
	\affiliation{Aix-Marseille Univ, CNRS, CNES, LAM, Marseille, France}
	
	
	
	\begin{abstract}
		Galaxy intrinsic alignment (IA) is both a source of systematic contamination to cosmic shear measurement and its cosmological applications, and a source of valuable information on the large scale structure of the universe and galaxy formation. The self-calibration (SC) method \citep{SC2008} was designed to separate IA from cosmic shear,  free of IA modeling. It was first successfully applied to the KiDS450 and KV450 data \citep{Yao2019}. We improved the SC method in several aspects, and apply it to the DECaLS DR3 shear + photo-z catalog and significantly improve the IA detection to $\sim 14\sigma$.  We find a strong dependence of IA on galaxy color, with strong IA signal ($\sim17.6\sigma$) for red galaxies, while the IA signal for blue galaxies is consistent with zero. The detected IA for red galaxies are in reasonable agreement with the non-linear tidal alignment model and the inferred IA amplitude increases with redshift. Our measurements rule out the constant IA amplitude assumption at $\sim3.9\sigma$ for the red sample. We address the systematics in the SC method carefully and performed several sanity checks. We discuss various caveats such as redshift/shear calibrations and possible improvements in the measurement,
		theory and parameter fitting that will be addressed in future works. 
	\end{abstract}
	
	\keywords{cosmology, gravitational lensing: weak, observations, large-scale structure of the universe, galaxy}
	
	
	% ---------------------------------
	\section{Introduction}
	For many cosmological probes, systematic errors in either observation
	or theory or both are becoming the dominant source of errors. They
	may already be responsible for several tensions in cosmology, such as
	the $H_0$ tension
	\citep{Riess2019,Planck2018I,Bernal2016,Lin2019,Freedman2019}. Another
	example is the $S_8=\sigma_8(\Omega_m/0.3)^{\alpha\sim 0.5}$ tension,
	between the Planck CMB experiment \citep{Planck2018I} and the stage
	III weak lensing surveys such as KiDS (Kilo Degree Survey,
	\cite{Hildebrandt2016,Hildebrandt2018,Asgari2020}), HSC (Hyper Suprime-Cam,
	\cite{HSC_Hamana2019,HSC_Hikage2019}), and DES (Dark Energy Survey,
	\cite{Troxel2017}), with the $S_8$ differences varies in between $\sim3\sigma$ and $\sim1\sigma$. A variety of tests have been
	carried out in investigating the $S_8$ tension (e.g. \citet{Asgari2019,Troxel2018,Chang2019,Joudaki2019}).
	
	Among systematic errors in weak lensing cosmology based on cosmic
	shear measurement, the galaxy intrinsic alignment (IA) is a prominent
	one. Cosmic shear is extracted from  galaxy shapes, with the underlying assumption that the intrinsic galaxy shapes have no spatial
	correlation. However, this assumption is invalid, since the large
	scale structure environment induces spatial correlation in the galaxy
	shapes. In the context of weak lensing, the spatially correlated part
	in the galaxy shapes (ellipticities) is called IA. It has been
	predicted by theory/simulations (e.g. \citet{Croft2000,Catelan2001,Crittenden2001,Jing2002,Hirata2004,Joachimi2013,Kiessling2015,Blazek2015,Blazek2017,Chisari2017,Xia2017}), and
	detected in observations (e.g.
	\citet{Lee2001,Heymans2004,BridleKing,Okumura2009,Dossett2013,Rong2015,Krause2016,Kirk2015,Troxel2017,Samuroff2019,Yao2019}. It
	is one of the key limiting factors to fully realize the power of weak
	lensing cosmology \citep{Heavens2002,Refregier2003,Hoekstra2008,LSST2009,Weinberg2013,TroxelIshak,Joachimi2015,Kilbinger2015,Mandelbaum2018}. 
	
	In cosmic shear data analysis, IA is often mitigated by fitting against an assumed fiducial IA template \citep{Troxel2017,Hildebrandt2016,Hildebrandt2018,HSC_Hamana2019,HSC_Hikage2019}. In
	contrast, the Self-Calibration (SC) methods \citep{SC2008,Zhang2010}
	were designed to remove the IA contamination without assumption on the
	IA model. This model independence is achieved, due to an intrinsic difference between the weak lensing field and the intrinsic alignment field. The former is a 2D (projected) field with a profound source-lens asymmetry, while the later is a
	statistically isotropic 3D field. The SC2008 method \citep{SC2008} has
	been applied to stage IV survey forecasts \citep{Yao2017,Yao2018},
	while the SC2010 method \cite{Zhang2010} has been examined in
	simulation \citep{Meng2018} and combined with SC2008 in the forecast
	\citep{Yao2018}. These studies showed that the SC method is generally accurate in IA removal/measurement. 
	
	\citet{Yao2019} first applied the SC2008 method to KiDS450
	\citep{Hildebrandt2016} and KV450 \citep{Hildebrandt2018} shear
	catalogs. To implement the SC method and to incorporate with various
	observational effects such as photo-z errors, \citet{Yao2019} built a
	Lensing-IA Separation (LIS) pipeline, and succeeded in the IA detection.  To further test the applicability of
	the SC method, and to improve the IA detection and applications, we
	apply the same LIS pipeline to the DECaLS (Dark Energy Camera Legacy
	Survey) DR3 shear catalog \citep{Phriksee2019}. Comparing to the previous work, we have significantly more galaxies and larger sky coverage. We use the photo-z obtained from k-nearest-neighbours
	\citep{Zou2019}. These improvements result in more significant IA
	detection, and allow us to reveal more detailed information on IA such as its redshift and color dependence. 
	
	This paper is organized as follows.  In \S \ref{Section method}, we
	briefly describe the SC method and the LIS pipeline. We also describe
	the theoretical model to compare with. \S \ref{Section data} describes
	the DECalS DR3 data used for the analysis. \S \ref{Section results}
	presents the main results and \S \ref{Section discussion} discusses
	further implications and possible caveats. We include more technical details in
	the appendix. 
	
	% -----------------------------------
	\section{The SC method and the LIS pipeline} \label{Section method}
	
	The observed galaxy shape $\gamma^{\rm
		obs}$ contains three components, 
	\ba
	\gamma^{\rm
		obs}=\gamma^{G}+\gamma^{N}+\gamma^{I}\ .
	\ea
	Here the superscript ``G'' denotes gravitational (G) lensing. The
	galaxy shape noise has a spatially uncorrelated part which we denote
	with the superscript ``N'',  and a spatially correlated part (the intrinsic
	alignment) which we denote with the superscript ``I''. When
	cross-correlating $\gamma^{\rm obs}$ with galaxy
	number density $\delta_g$, the  $\gamma^{N}$ term has no
	contribution.  The measured correlation will contain two parts, 
	\begin{equation}
	\langle \gamma^{\rm obs}\delta_g \rangle ~=~ \langle \gamma^G
	\delta_g \rangle + \langle \gamma^I \delta_g \rangle \ . \label{Eq gamma-g corr}
	\end{equation}
	The first term on the right-hand side of the equation is the (lensing
	part) Gg correlation, and the second term is the (IA part) Ig
	correlation.  The first step of SC2008 is to separate and measure Ig (and
	Gg), without resorting to IA modeling. The second step is to convert Ig into the GI
	term contaminating the measurement of cosmic shear auto-correlation, through a scaling relation found in \citet{SC2008}. The current paper is restricted to the first step, since no results on the cosmic shear auto-correlation will be presented here. We focus on the
	Ig measurement and its application. 
	
	\subsection{Separating Gg and Ig} \label{Section Ig}
	For a pair of galaxies, we denote the photo-z of the galaxy used for
	shape measurement as $z_{\gamma}^P$, and the photo-z of the galaxy
	used for number density measurement as $z_g^P$. Both the intrinsic
	alignment and the galaxy number density fields are statistically
	isotropic 3D fields. Therefore  the $\langle Ig \rangle$ correlation with
	$z_{\gamma}^P<z_g^P$  is identical to $\langle Ig \rangle$ with
	$z_{\gamma}^P>z_g^P$. Namely, it is insensitive to the ordering of
	$(z^P_\gamma, z_g^P)$ pair in redshift space. This holds for both real (spectroscopic)
	redshift and photometric redshift. In contrast, the lensing correlation
	requires $z_\gamma>z_g$ for the true redshift (z).  Therefore in the
	photo-z ($z^P$) space, the $\langle Gg \rangle$ correlation is
	smaller for the pairs with $z_{\gamma}^P<z_g^P$,
	compared with the $z_{\gamma}^P > z_g^P$ pairs.\footnote{In the limit of
		negligible photo-z error, the $\langle Gg \rangle$ correlation
		vanishes for  $z_{\gamma}^P<z_g^P$ pairs. In reality, photo-z has both
		scatters and outliers, the $\langle Gg \rangle$ correlation persists
		even for  $z_{\gamma}^P<z_g^P$ pairs.} 
	
	Therefore we can form two sets of two-point statistics measured
	from the same data in the same photo-z bin (e.g. the $i$-$th$ photo-z
	bin). In terms of the angular power spectrum,  
	\begin{subequations}
		\begin{align} 
		C^{\gamma g}_{ii}&=C^{Gg}_{ii}+C^{Ig}_{ii}, \label{Eq gamma-g} \\
		C^{\gamma g}_{ii}|_S&=C^{Gg}_{ii}|_S+C^{Ig}_{ii} \label{Eq
			gamma-g|S}\ .
		\end{align}
	\end{subequations}
	Here $C^{\gamma g}_{ii}$ is the galaxy shape-number density angular
	power spectrum for all pairs in the $i$-th redshift bin, while $C^{\gamma g}_{ii}|_S$ is the one
	only for pairs with $z^P_\gamma<z^P_g$. According to the above
	analysis, with this ``$|_S$'' selection, the lensing signal drops
	from $C^{Gg}_{ii}$ to $C^{Gg}_{ii}|_S$, while the IA signal
	$C^{Ig}_{ii}$ remains the same. 
	
	The drop in the lensing signal can be determined  by the $Q$ parameter, 
	\begin{equation} \label{Eq Q}
	Q_i(\ell)\equiv \frac{C^{Gg}_{ii}|_S(\ell)}{C^{Gg}_{ii}(\ell)}\ .
	\end{equation}
	$Q(\ell)$ has only weak dependence on cosmology and
	$\ell$ \citep{SC2008,Yao2017}. This makes the SC method cosmology-independent
	to good accuracy. But it is
	sensitive to the photo-z quality. $Q=0$ for perfect photo-z (photo-z is accurate so that lensing signal drops fully due to the selection), 
	$Q\rightarrow 1$ for poor photo-z, and $Q\in
	(0,1)$ in general. We are then able to separate Gg and Ig \citep{SC2008,Yao2019}, 
	\begin{align} 
	C^{Gg}_{ii}(\ell)&=\frac{C^{\gamma g}_{ii}(\ell)-C^{\gamma g}_{ii}|_S(\ell)}{1-Q_i(\ell)}, \label{Eq SC C^Gg}\\
	C^{Ig}_{ii}(\ell)&=\frac{C^{\gamma g}_{ii}|_S(\ell)-Q_i(\ell)C^{\gamma g}_{ii}(\ell)}{1-Q_i(\ell)}. \label{Eq SC C^Ig}
	\end{align}
	
	In this work, we extended the formalism of SC to the correlation function, considering two additional effects comparing to previous works \cite{SC2008,Yao2017,Yao2019}: (1) the scale-dependent $Q_i(\theta)$ and (2) impact from non-symmetric redshift distribution, leading to $w^{Ig}_{ii}|_S\ne w^{Ig}_{ii}$, or $C^{Ig}_{ii}|_S\ne C^{Ig}_{ii}$. As a result, we have
	\begin{subequations}
		\begin{align} 
		w^{\gamma g}_{ii}(\theta)&=w^{Gg}_{ii}(\theta)+w^{Ig}_{ii}(\theta), \label{Eq wgamma-g} \\
		w^{\gamma g}_{ii}|_S (\theta)&=w^{Gg}_{ii}|_S (\theta)+w^{Ig}_{ii}|_S(\theta) \label{Eq
			wgamma-g|S}\ ,
		\end{align}
	\end{subequations}
	which give us 
	\begin{subequations}
		\begin{align}
		\label{Eq Gg correlation}
		w^{Gg}_{ii}(\theta)&=\frac{Q^{Ig}_{i}(\theta)w^{\gamma g}_{ii}(\theta) - w^{\gamma g}_{ii}|_S(\theta)}{Q^{Ig}_i(\theta)-Q^{Gg}_{i}(\theta)},\\
		\label{Eq Ig correlation}
		w^{Ig}_{ii}(\theta)&=\frac{w^{\gamma g}_{ii}|_S(\theta) -Q^{Gg}_{i}(\theta) w^{\gamma
				g}_{ii}(\theta)}{Q^{Ig}_i(\theta)-Q^{Gg}_{i}(\theta)}\ .
		\end{align}
	\end{subequations}
	Here the $Q$ values are calculated theoretically with a fiducial cosmology and the redshift distributions from data. $Q^{Gg}_i$ is defined as
	\begin{equation} \label{Eq Q^Gg(theta)}
	Q^{Gg}_i(\theta) \equiv w^{Gg}_{ii}|_S(\theta)/w^{Gg}_{ii}(\theta),
	\end{equation}
	which is similar as the previous definition Eq.\,\eqref{Eq Q} using angular power spectra. With this definition, we no longer need to assume a constant $\bar{Q}_i$ value as before \citep{Yao2019}, instead, the angular scale dependency $Q^{Gg}_i(\theta)$ is taken into consideration, for a more precise lensing-IA separation.
	
	Similarly, $Q^{Ig}_i$ is defined as
	\begin{equation} \label{Eq Q^Ig(theta)}
	Q^{Ig}_i(\theta) \equiv w^{Ig}_{ii}|_S(\theta)/w^{Ig}_{ii}(\theta)
	\end{equation}
	to account for the non-symmetric redshift distribution, which could potentially make $Q^{Ig}_i$ deviates from 1 ($w^{Ig}_{ii}|_S\ne w^{Ig}_{ii}$).
	
	Here \{ $w^{\gamma g}$, $w^{\gamma g}|_S$ \} are direct
	observables and \{ $Q^{Gg}_i$, $Q^{Ig}_i$ \} can be robustly calculated given photo-z PDF, so
	we are able to separate and measure both $w^{Gg}$ and $w^{Ig}$ as in Eq.\,\eqref{Eq Gg correlation} and \eqref{Eq Ig correlation}.  A key
	step in our method is to calculate $Q$. The calculation is
	straightforward, but technical. We present detailed description in
	the appendix.

	\subsection{Interpreting the separated Gg and Ig} \label{Section
		theory}
	The next step is to extract the physics out of the Gg and Ig separated above.  We need to compare with the theoretically predicted $w^{Gg}$ and
	$w^{Ig}$.  In this section, we briefly describe the basic theory of weak lensing and intrinsic alignment.  The comparison between theory and observation will be presented in \S \ref{Section results}. 
	
	The lensing-galaxy cross power spectrum is calculated by the Limber
	equation, 
	\ba
	C^{Gg}_{ii}(\ell)=\int_0^\infty\frac{W_i(\chi)n_i(\chi)}{\chi^2}
	b_g
	P_\delta\left(k=\frac{\ell}{\chi};\chi\right)d\chi\
	.
	\label{Eq Gg}
	\ea
	Here $W_i$ is the lensing efficiency function. For a flat universe, 
	\begin{equation} \label{Eq W_i}
	W_i(\chi_L)=\frac{3}{2}\Omega_m\frac{H_0^2}{c^2}(1+z_L)
	\int_{\chi_L}^\infty
	n_i(\chi_S)\frac{(\chi_S-\chi_L)\chi_L}{\chi_S}d\chi_S \ .
	\end{equation}
	$n_i(\chi)$ is the galaxy distribution of the $i^{\text{th}}$ photo-z
	bin in the comoving distance space, and is linked to the galaxy
	distribution in the true redshift space by
	$n_i(\chi)=n_i(z) dz/d\chi$. Here $\chi$ is the comoving
	distance, $b_g$ is the galaxy bias, and $P_\delta$ is the matter power
	spectrum.
	Similarly, the IA-galaxy cross angular power spectrum $C^{Ig}$ is
	given by
	\begin{equation}
	C^{Ig}_{ii}(\ell)=\int_0^\infty\frac{n_i(\chi)n_i(\chi)}{\chi^2}b_gP_{\delta,\gamma^I}\left(k=\frac{\ell}{\chi};\chi\right)d\chi. \label{Eq Ig}
	\end{equation}
	In this expression, $P_{\delta,\gamma^I}$ is the 3D matter-IA power
	spectrum, which depends on the IA model being used (or the ``true'' IA
	model). For comparison,  we adopt  the non-linear tidal alignment
	model \citep{Catelan2001,Hirata2004} as the fiducial IA
	model. It  is widely used in the other
	stage III surveys
	\citep{Hildebrandt2016,Hildebrandt2018,Troxel2017,HSC_Hikage2019,HSC_Hamana2019,Chang2019}. In
	this model, 
	\begin{equation} \label{Eq IA 3D}
	P_{\delta,\gamma^I}=-A_{\rm IA}(L,z)\frac{C_1\rho_{m,0}}{D(z)}P_\delta(k;\chi),
	\end{equation}
	where $\rho_{m,0}=\rho_{crit}\Omega_{m,0}$ is the mean matter density
	of the universe at $z=0$. $C_1=5\times 10^{-14}(h^2M_{\rm sun}/{\rm
		Mpc}^{-3})$ is the empirical amplitude found in
	\cite{BridleKing}. In this work we adopt $C_1\rho_{crit}\approx
	0.0134$ as in \cite{Krause2016,Yao2019}. $D(z)$ is the linear growth
	factor normalized to 1 today. $A_{\rm IA}(L,z)$ is the IA amplitude
	parameter, which is expected to be luminosity($L$)- and
	redshift($z$)-dependent. In this work, we will investigate the possible
	redshift dependence and the galaxy-type dependence of this $A_{\rm IA}$ parameter.

	The theoretical prediction of $w^{Gg}$ and $w^{Ig}$ are then given by
	the Hankel transformation, 
	\begin{equation} \label{Eq Hankel}
	w(\theta)=\frac{1}{2\pi}\int d\ell~\ell C(\ell) J_2(\ell\theta)\ .
	\end{equation}
	Here $J_2(x)$ is the Bessel function of the first kind of order 2.  We
	adopt the CCL library \footnote{Core Cosmology Library,
		\url{https://github.com/LSSTDESC/CCL}} \citep{Chisari2019CCL} for
	the theoretical calculations. These results are cross-checked with
	CAMB \footnote{Code for Anisotropies in the Microwave Background,
		\url{https://camb.info/}} \citep{Lewis2000CAMB} in previous work
	\citep{Yao2019}. The cosmological parameters being used to calculate
	the theoretical predictions are the best-fit cosmology of Planck2018 and KV450, as
	shown in Table\,\ref{table: fiducial cosmology}. The
	impact from uncertainties in the cosmological parameters on the
	theoretical predictions is negligible, compared with that from
	uncertainties in the  galaxy bias $b_g$ and the IA amplitude $A_{\rm
		IA}$. Also, $\sigma_8$ strongly degenerates with $b_g$ in our case and they both enter the estimation of $w^{Gg}$ and $w^{Ig}$ in the
	same way. Therefore for the purpose of studying IA, it is valid to fix
	the cosmology. 
	
	\begin{table}
		\centering
		\caption{The $\Lambda$CDM cosmological parameters adopted in our analysis, which correspond to the best-fit cosmology from Planck2018 \cite{Planck2018I} (fiducial) and KV450 \citep{Hildebrandt2018} (alternative). \label{table: fiducial cosmology}}
		%     \begin{ruledtabular}
		\begin{tabular}{ c c c c c c c }
			\hline
			Survey & $h_0$ & $\Omega_b h^2$ & $\Omega_c h^2$ & $n_s$ & $\sigma_8$ & $w$  \\
			\hline
			Planck & 0.6732  &  0.022383  &  0.12011  &  0.96605  & 0.812 &  -1.0 \\
			KV450 & 0.745  &  0.022  &  0.118  &  1.021  & 0.836 &  -1.0 \\
			\hline
		\end{tabular}
		%     \end{ruledtabular}
	\end{table}
	
	% -------------------------------------------------------------
	
	\section{Survey data} \label{Section data}
	
	%DECaLS (Dark Energy Camera Legacy Survey) DR3 shear catalog \citep{Phriksee2019}, with photo-z obtained from k-nearest-neighbours \citep{Zou2019}

	% ------- shear --------
	We apply our method to  the Dark Energy Camera Legacy Survey (DECaLS)
	Data Release 3, which is part of the Dark Energy Spectroscopic
	Instrument (DESI) Legacy Imaging Surveys \citep{Dey2019}.  The DECaLS
	DR3 contains images covering 4300 $\rm deg^2$ in g-band, 4600 $\rm
	deg^2$  in 
	r-band and 8100 $\rm deg^2$ in z-band. In total 4200 $\rm deg^2$ have been observed in all three optical bands. The DECaLS data are processed by Tractor \citep{Meisner2017,Lang2014}.
	
	The sources from the Tractor catalog are divided into five
	morphological types. Namely, 
	\begin{enumerate}
		\item Point sources (PSF),
		\item Simple  galaxies (SIMP: an exponential profile with a fixed 0.45" effective
		radius and round profile), 
		\item de Vaucouleurs  (DEV: elliptical galaxies), 
		\item Exponential (EXP: spiral galaxies), 
		\item Composite model (COMP: composite profiles 
		which are de Vaucouleurs and exponential with the same source center).
	\end{enumerate}
	In this catalog, the sky-subtracted images are stacked in five 
	different ways: one stack per band, one “flat” Spectral Energy Distribution (SED) stack of each g-, r- and z-band, one “red” (g-r=1 mag and r-z=1 mag)
	SED stack of all bands. The sources are kept above the detection limit in any stack as 
	candidates. The PSF model (delta function) and the SIMP model are adjusted on individual images, which are convolved by their own 
	PSF model.
	
	The galaxy ellipticities $e_{1,2}$ are free parameters of the above four SIMP, DEV, EXP and COMP models, except for the PSF model. 
	The ellipticity are estimated by a joint fit on the three optical g-, r-, and z-band. We model potential measurement bias with a 
	multiplicative ($m$) and additive bias ($c$) \citep{Heymans2012,Miller2013, Hildebrandt2016},
	\begin{equation} \label{Eq shape correction}
	\gamma^{\rm obs}=(1+m)\gamma^{\rm true}+c,
	\end{equation}
	The additive bias  is expected to come from residuals in the
	anisotropic PSF correction. It depends on galaxy sizes. The addtive bias $c$ is subtracted from each galaxy in the catalog. The 
	multiplicative bias comes from the shear measurement. It can be generated
	by many effects, such as measurement  
	method \citep{Mandelbaum2015}, blending and crowding \citep{Martinet2019}. In order to calibrate our shear catalog, 
	we cross-matched the DECaLS DR3 objects with the Canada-France-Hawaii
	Telescope (CFHT) Stripe 82 objects, and then computed the correction
	parameters \citep{Phriksee2019}. In addition, the data from DECaLS DR3
	catalog were tested with  
	the Obiwan simulations (Burleigh et al. in prep., \cite{Kong2020}), also described in Table A1 in \cite{Phriksee2019}. 
	% -------------------------------

	\begin{figure}\centering
		\includegraphics[width=1.1\columnwidth]{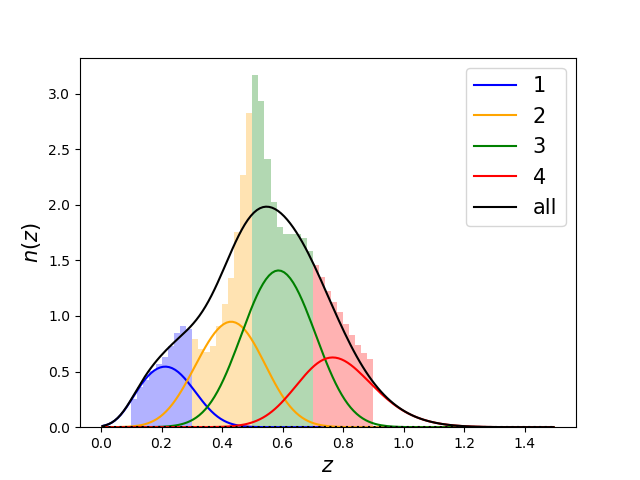}
		\caption{The redshift distribution of the galaxy samples
			analyzed. The shaded histogram is the photo-z distribution, which is divided into 4 tomographic bins. The color curves are the estimated true redshift distributions $n_i(z)$, while the black curve gives the total $n(z)$.}
		\label{fig: n(z)}
	\end{figure}

	% ------- photo-z------------
	We employ the photo-z from \cite{Zou2019}, which is based on the algorithm of k-nearest-neighbors and local linear regression. The photo-z is obtained from 5 photometric bands: three optical bands (g, r, and z), and two infrared bands (Wide-field Infrared Survey Explorer W1 and W2). We use samples with $r<23$ mag. The training sample includes $\sim2.2$ M spectroscopic galaxies.
	% -------------------------------
	
	For each galaxy we use in this work, we add two extra selections. One is to remove some galaxies with extreme shear multiplicative bias (requiring $1+m>0.5$). We note many selection effects could potentially bias the shear calibration, with more details in \cite{Li2020,Huff2017,Sheldon2017} and its potential impact in this work in Appendix \ref{Apdx: m and n(z) bias}. The other is requiring small estimated photo-z error ($\Delta_z^P<0.1$). Together with the selection of $0.1<z^P<0.9$, we obtain 23 million galaxies for the SC
	analysis. We divide them into 4 photo-z bins  ($0.1<z^P<0.3$,
	$0.3<z^P<0.5$, $0.5<z^P<0.7$ and $0.7<z^P<0.9$). For each galaxy, our
	kNN photo-z algorithm also provides an Gaussian estimation of the photo-z
	error. We further apply this Gaussian scatter to obtain the redshift probability distribution function (PDF) for each galaxy. The overall photo-z distribution $n_i^P(z^P)$ and the true-z
	distribution $n_i(z)$ are shown in
	Fig.\,\ref{fig: n(z)}. More detailed discussion on the photo-z quality are included in Appendix \ref{Apdx: photo-z}, where we show the Gaussian PDF is not accurate, however the overall scatter is accurate, which is of most importance in the self-calibration analysis \cite{Zhang2010}. The possible impact of biased $n(z)$ is discussed in Appendix \ref{Apdx: m and n(z) bias}.
	
	\section{Results} \label{Section results}
	We present the measurement of $w^{\gamma g}$ and $w^{\gamma g}|_S$ in \S \ref{Section
		results observables},  $Q^{Gg}$ and $Q^{Ig}$ in \S \ref{Section
		results Q},  $w^{Gg}$
	and $w^{Ig}$ in \S \ref{Section results lensing-IA}. All the analysis in this work uses the default pipeline developed by JY in
	\cite{Yao2019}. The 2-point correlation functions described in
	Eq.\,\eqref{Eq gamma-g estimator} is performed with
	TreeCorr\footnote{\url{https://github.com/rmjarvis/TreeCorr}} code
	\citep{Jarvis2004}.
	
	\subsection{$w^{\gamma g}$ and $w^{\gamma g}|_S$ measurement} \label{Section results observables} 
	
	We adopt the following estimator
	\citep{Mandelbaum2006,Singh2017,Yao2019} to calculate $w^{\gamma g}$
	and $w^{\gamma g}|_S$, 
	\begin{equation} \label{Eq gamma-g estimator}
	w^{\gamma g}=\frac{\sum_{\rm ED}w_j\gamma^+_j}{\sum_{\rm
			ED}(1+m_j)w_j}-\frac{\sum_{\rm ER}w_j\gamma^+_j}{\sum_{\rm
			ER}(1+m_j)w_j}\ .
	\end{equation}
	Here $\sum_{\rm ED}$ means summing over all the tangential ellipticity
	(E) - galaxy number counts in the data (D) pairs, $\sum_{\rm ER}$
	means summing over all the tangential ellipticity (E) - galaxy number
	counts in the random catalog (R) pairs. The numerators give the
	stacked tangential shear weighted by the weight $w_j$ from the shear
	measurement algorithm of the $j^{\rm th}$ galaxy. The denominators
	give the normalization considering the number of pairs, the shear
	weight $w_j$, and the calibration for shear multiplicative bias
	$(1+m_j)$. Here we note that, after normalization with the number of galaxies,
	the two denominators $\sum_{\rm ED}(1+m_j)w_j$ and $\sum_{\rm
		ER}(1+m_j)w_j$ are generally considered the same at large scale
	of our interest,  as the boost factor (the ratio of these two) is
	normally considered as 1
	\citep{Mandelbaum2005boostfactor,Singh2017}.

	\begin{figure}\centering
		\hspace*{-0.5cm}
		\includegraphics[width=1.1\columnwidth]{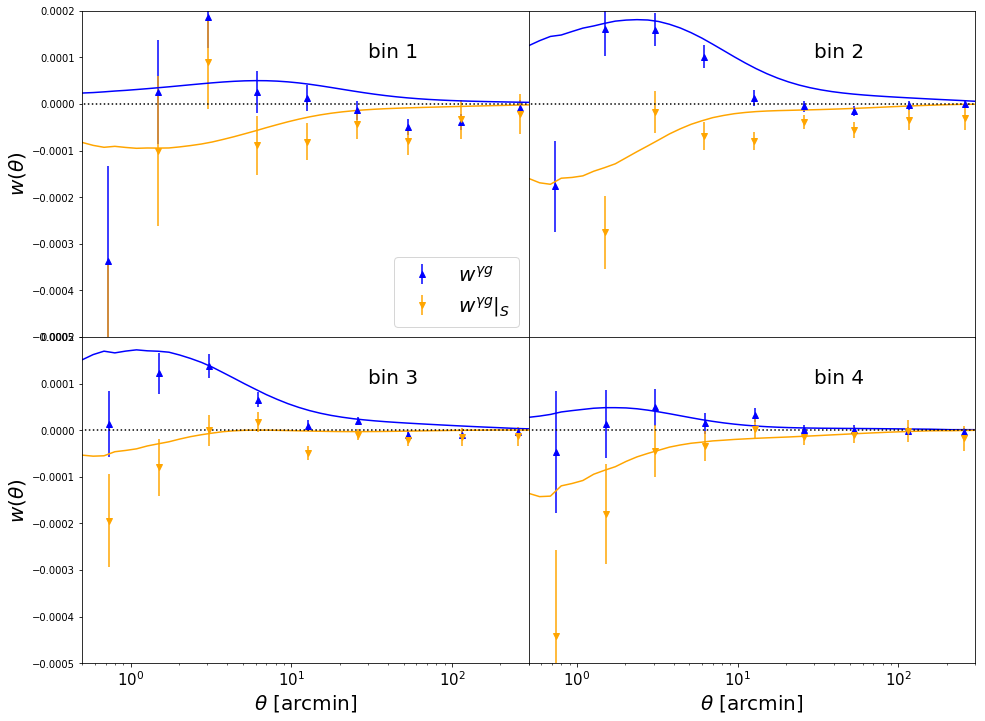}
		\caption{The directly measured $w^{\gamma g}$ (blue up-triangles) and $w^{\gamma g}|_S$ (orange
			down-triangles), along with the theory curves. The  pair weighting
			adopted in $w^{\gamma g}|_S$  mainly down-weights the 
			lensing contribution, while the IA contribution is almost
			unchanged. The difference between the two then quantifies the
			efficiency of the SC method. The difference is statistically significant in
			all 4 redshift bins ($5.7\sigma$, $16.1\sigma$, $10.6\sigma$,
			$6.0\sigma$). We note that the theoretical curves are not the
			best-fit for $w^{\gamma g}$ and $w^{\gamma g}|_S$, but what predicted from the best-fit of
			separated signals $w^{\rm Gg}$ and $w^{Ig}$ in Fig.\,\ref{fig: lensing IA separation all} with the scale-dependent $Q^{Gg}_i(\theta)$ and $Q^{Ig}_i(\theta)$ in Fig.\,\ref{fig: Q} and \ref{fig: Q_Ig}. The fitting $\chi^2$ are [31.9, 65.0, 40.8, 8.4] with $d.o.f.=16$ for each bin. There is a visual mismatch that is partially due to the strong correlation shown in Fig.\,\ref{fig: r_obs all}.}
		\label{fig: observables}
	\end{figure}
	
	For the random catalog, we use the DECaLS DR7 random catalog\footnote{\url{http://legacysurvey.org/dr7/files/}} and fit it into the DECaLS DR3 shear catalog footprint \citep{Phriksee2019} with Healpy\footnote{\url{https://github.com/healpy/healpy}}. The size of our random catalog is $\sim10$ times the size of the whole DECaLS DR3 shear catalog. This random catalog is used in Eq.\,\eqref{Eq gamma-g estimator} for the ``R'' part, while for the ``D'' part we use the galaxies in each tomographic bin. So the random sample size is much larger than real data. After the random-subtraction, the null-test with $\gamma^X$ (the $45\deg$ rotation of $\gamma^+$) of Eq.\,\eqref{Eq gamma-g estimator} is consistent with zero.
	
	We note that we are not including the sky varying survey depth in
	the random sample, for three reasons. 
	(1) Since our photo-z sample has a cut with $r<23$ \citep{Zou2019} to maintain high galaxy completeness, the ``fake overdensity'' due to this effect is expected to be low \citep{Raichoor2017}.
	(2) The small (due to the previous point)``fake overdensity'' from varying observational depth is
	expected to not correlate with the galaxy shapes, as both the
	lensing part and the IA part are parts of the large scale structure.
	Therefore the fractional contribution in the correlations as selection bias is expected to be even less than in the density field.
	(3) Even if there still exists a selection bias in the 2-point
	statistics, it should be captured by our Jackknife re-sampling and is
	therefore appropriately included in the covariance matrix. 
	In the next generation surveys, more detailed consideration for the random catalog should also be addressed.
	
	\begin{figure}\centering
		\hspace*{-0.3cm}
		\includegraphics[width=1.1\columnwidth]{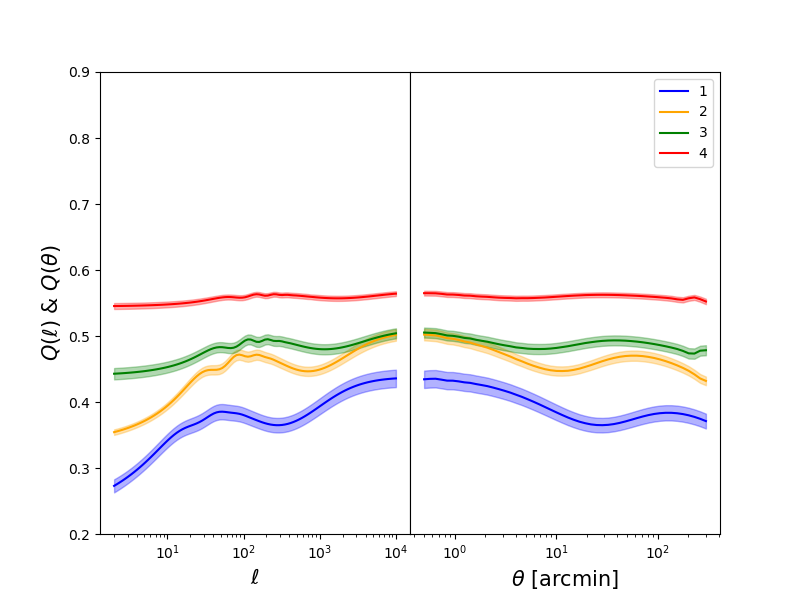}
		\caption{We show the measured $Q^{Gg}_i(\ell)$ from power spectra (as in Eq.\,\eqref{Eq Q}) in the left panel and $Q^{Gg}_i(\theta)$ from correlation functions (as in Eq.\,\eqref{Eq Q^Gg(theta)}) in the right panel. Different colors represent different bins. The shaded area shows 20 times the statistical error on the $Q$ values. In the right panel we show the angular range $0.5<\theta<300$ [arcmin] that we are interested in, before any angular cut being adopted.} 
		\label{fig: Q}
	\end{figure}
	
	\begin{figure}\centering
		\hspace*{-0.3cm}
		\includegraphics[width=1.1\columnwidth]{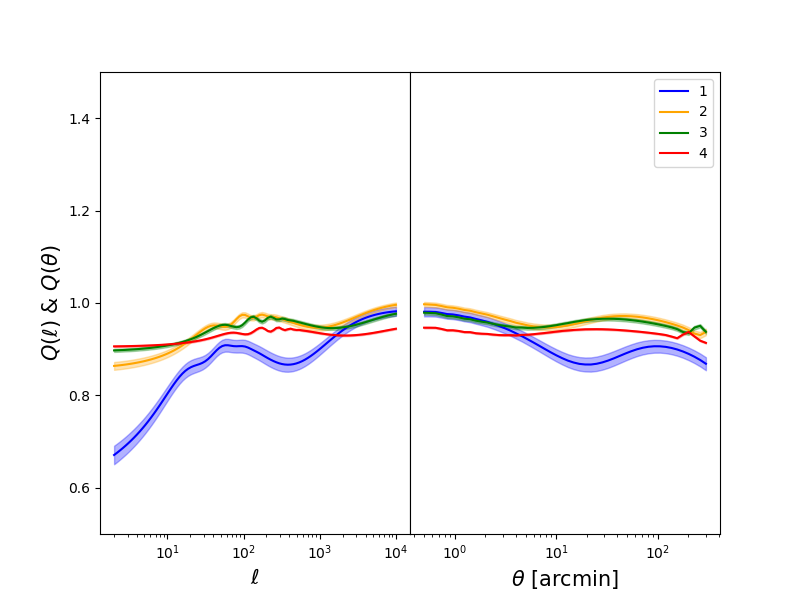}
		\caption{Similar to Fig.\,\ref{fig: Q}, but for $Q^{Ig}_i(\ell)$ and $Q^{Ig}_i(\theta)$ (as in Eq.\,\eqref{Eq Q^Ig(theta)}). The small deviation of $Q^{Ig}$ from 1 ($\sim 10\%$ level) comes from the non-symmetric distribution of $n^P(z^P)$ and $n(z)$, see Fig.\,\ref{fig: n(z)} for example. Ignoring this will cause a $\sim 20\%$ bias in $w^{Ig}$ measurement. More discussions are included in the main text.} 
		\label{fig: Q_Ig}
	\end{figure}

	We use Jackknife re-sampling to obtain the covariance matrices of
	$w^{\gamma g}$, $w^{\gamma g}|_S$, $Q^{Gg}_i$, $Q^{Ig}_i$, and the derived $w^{Gg}$ and
	$w^{Ig}$. We use a K-means clustering code kmeans\_radec\footnote{\url{https://github.com/esheldon/kmeans_radec}} and generate 500 Jackknife regions. The choice of 500 Jackknife regions is to prevent biased estimation of the covariance for the length 34 data vector we are going to use (discussed in \S \ref{Section results lensing-IA}), based on the analysis of \cite{Mandelbaum2006,Hartlap2007}.
	
	Fig.\,\ref{fig: observables} shows the measured $w^{\gamma g}$ and
	$w^{\gamma g}|_S$. The observed $w^{\gamma g}$ and
	$w^{\gamma g}|_S$ at all four redshift bins are statistically different, with $5.7-16.1\sigma$
	significance. It suggests that the photo-z quality sufficient for our need, and the selection $z_\gamma^P<z_g^P$ is efficient to reduce the lensing contribution, This clear separation is a necessary condition for our SC method.
	
	The  $w^{\gamma g}$-$w^{\gamma g}|_S$ separation is clearly more significant in this work than in
	\cite{Yao2019}, which used KiDS450 and KV450 data. This we think is mainly due to the larger galaxy number in our DECaLS sample,
	especially in the second and the third redshift bins. Differences in the photo-z algorithm adopted and the resulting photo-z quality may also matter. However, since we lack robust information on photo-z
	outliers to quantify its impact on SC, we leave this issue for further study. 
	
	We also show the theoretical curves in Fig.\,\ref{fig:
		observables} and calculated how good those fitting $\chi^2$ are comparing to data. This demonstrates that the nonlinear tidal alignment model can provide a reasonably good description of the measurement. 
	
	Nevertheless, we caution that they are not the best-fit for 
	$w^{\gamma g}$ and $w^{\gamma g}|_S$, but the prediction from the best-fit for $w^{Gg}$ and $w^{Ig}$, which we will discuss in the
	next subsection. The two data sets (\{$w^{\gamma g}$, $w^{\gamma g}|_S$\} v.s. \{$w^{Gg}$, $w^{Ig}$\}) are identical if we have perfect
	knowledge of $Q^{Gg}$ and $Q^{Ig}$. In this work, we choose to fit against $w^{Gg}$ and $w^{Ig}$,
	since  their physical meanings (the lensing-galaxy correlation and the
	IA-galaxy correlation) are more straightforward, compared with
	$w^{\gamma g}$ and $w^{\gamma g}|_S$. The
	reasonably good agreement (Fig.\,\ref{fig: observables}) show that,
	our best-fit with scale cuts for \{$w^{Gg}$, $w^{Ig}$\} also agrees very well with the \{$w^{\gamma g}$, $w^{\gamma g}|_S$\} measurements. In the future analysis, we can alternatively use $w^{\gamma g}$ and $w^{\gamma
		g}|_S$ directly for the fitting.  For such exercise, we also need
	the covariance matrix of the two sets of observables.  We discuss them in
	in the Appendix \ref{Apdx: cov observable} and Fig.\,\ref{fig: r_obs
		all} for your interests. As expected, the two have a strong positive
	correlation, since $w^{\gamma g}|_S$  is totally and positively
	included in $w^{\gamma g}$. Such a strong correlation must be taken into
	account in the related data analysis. Also due to this strong correlation, the fitted curves are visually different from data at some level, while the fitting $\chi^2$ are reasonable as shown in Fig.\,\ref{fig: observables}.
	
	\subsection{The lensing-drop $Q^{Gg}_i$ and IA-drop $Q^{Ig}_i$} \label{Section results Q}
	
	Fig.\,\ref{fig: Q}  shows the measured lensing-drop $Q^{Gg}_i(\ell)$ from power spectra definition Eq.\,\eqref{Eq Q} and $Q^{Gg}_i(\theta)$ from correlation function definition Eq.\,\eqref{Eq Q^Gg(theta)}. We leave calculation details in the appendix \ref{Apdx:
		eta}. As we have explained in \S \ref{Section Ig}, $Q^{Gg}$ is mainly
	determined by the photo-z quality, with $Q^{Gg}=0$ for perfect photo-z
	and $Q^{Gg}=1$ for totally wrong photo-z. For the SC method to be
	applicable, $Q^{Gg}$ must be significantly smaller than
	unity \citep{SC2008,Yao2019}. Fig.\,\ref{fig: Q} showed that $Q_i(\ell)\sim 0.5$ for a wide
	range of $\ell$ and photo-z bin. Therefore the photo-z quality is
	already sufficiently good to enable the SC method. $Q$ varies
	between photo-z bins. We tested that for photo-z outlier rate $<20\%$, the bias in $Q$ for the current stage surveys is negligable. Besides the difference in photo-z quality, the
	effective width of the lensing kernel ($W_L(z_S,z_L)$) also plays a
	role.

	According to Fig.\,\ref{fig: Q}, as well as in our previous work
	\citep{Yao2019}, the $Q^{Gg}_i$ value is roughly constant in the range of
	$50<\ell<3000$. This is the main regime of interest in weak lensing cosmology. Previously we adopted the approximation
	$\bar{Q}_i=\langle Q_i(\ell)\rangle$, which could potentially under-estimate the IA signal at small-scale and over-estimate the IA signal at large-scale. In this work, by using scale-dependent $Q(\theta)$, as shown in the right panel of Fig.\,\ref{fig: Q}, we get rid of this effect. However, we note that as photo-z quality improves and/or redshift increases, the Q value will become more scale-independent so the above approximation should still hold. Thus this is not a major problem, but still worth bringing out. 
	
	In Fig.\,\ref{fig: Q} we also include the statistical error. They are shown in the shaded regions, while the error-bars are exaggerated (20 times). The fact that the Q values have very low statistical error proves our previous statements in \cite{Yao2017,Yao2019}.
	
	Similarly, we show the $Q^{Ig}_i$ measurements in Fig.\,\ref{fig: Q_Ig}. Generally $Q^{Ig}_i\sim 1$ is a good assumption. However, due to the non-symmetric photo-z distribution $n^P_i(z^P)$ and true-z distribution $n_i(z)$ shown in Fig.\,\ref{fig: n(z)}, the $Q^{Ig}$ for real data will deviate from 1. We tested that for the $\sim 10\%$ over-estimation for $Q^{Ig}$ (if assumed to be 1) shown in Fig.\,\ref{fig: Q_Ig}, the resulting $w^{Ig}$ will be under-estimated by $\sim 20\%$. Interestingly, the final estimation of the IA amplitude $A_{\rm IA}$ is almost unbiased (see later in Fig.\,\ref{fig: bg-IA fit all QIg}), which is due to the corresponding changes in the covariance matrix as well as the $w^{Gg}$ signal.
	
	Furthermore, we tested how the $Q$ parameters depend on the assumed fiducial cosmology. We compared the calculation of $Q^{Gg}$ and $Q^{Ig}$ with Planck2018 cosmology and KV450 cosmology (where the main $S_8$ tension resides), as shown in Table \ref{table: fiducial cosmology}. The differences are at $\sim 10^{-3}$ to $\sim 10^{-5}$ level, and the resulting bias in $w^{Ig}$ is $\sim 10^{-3}$ level. This proved our previous statement in \cite{Yao2017,Yao2019} that, by construct, the $Q^{Gg}$ and $Q^{Ig}$ measurements are insensitive to the fiducial cosmology. For the same reason, $Q^{Ig}$ is also insensitive to the assumed IA model.
	
	\subsection{Lensing-IA Separation (LIS)} \label{Section results lensing-IA}
	With the measured \{ $w^{\gamma g}$, $w^{\gamma g}|_S$ \} (Fig.\ref{fig:
		observables}), $Q^{Gg}_i$ (Fig.\,\ref{fig: Q}) and $Q^{Ig}$ (Fig.\,\ref{fig: Q_Ig}), we are then able to
	separate $w^{Gg}$ and $w^{Ig}$ by Eq.\,\eqref{Eq Gg correlation} \&
	\eqref{Eq Ig correlation}. The results are shown in Fig.\,\ref{fig:
		lensing IA separation all}, along with the normalized covariane
	matrix (Fig. \ref{fig: r_GgIg all}). We cut off small-scales to prevent further contamination from non-linear galaxy bias, massive neutrinos, baryonic effects, boost factor, etc. We cut off large-scale to prevent impact from insufficient random catalog. The cuts are shown in the grey shaded regions. The detection of intrinsic alignment
	($w^{Ig}$) is significant at all four redshift bins and the
	corresponding S/N=$3.5$, $11.9$, $5.5$, $4.1$ respectively.\footnote{We caution
		that the detection significance is likely overestimated, since we do
		not include uncertainties in the $Q$ value. The induced fluctuation is
		$\delta w^{Ig}=-w^{Gg} \delta Q/(1-Q)\simeq -w^{Gg}\times (2\delta
		Q)$. Since $w^{Ig}\sim w^{Gg}$ for the full sample, the induced
		fractional error is $\delta w^{Ig}/w^{Ig}\sim -2\delta Q$. The
		statistical $Q$ fluctuation  estimated by the
		Jackknife method is $\sim 10^{-3}$, and is therefore  negligible in
		the $w^{Ig}$ error budget. However, systematic error of $Q$ arising
		from photo-z outliers may be larger. Unless $|\delta Q|\gtrsim 0.05$,
		the detection significance of $w^{Ig}$ will not be significantly
		affected. After we have reliable estimation on photo-z outliers, we
		will quantify its impact. }
	
	\begin{figure}\centering
		\hspace*{-0.5cm}
		\includegraphics[width=1.1\columnwidth]{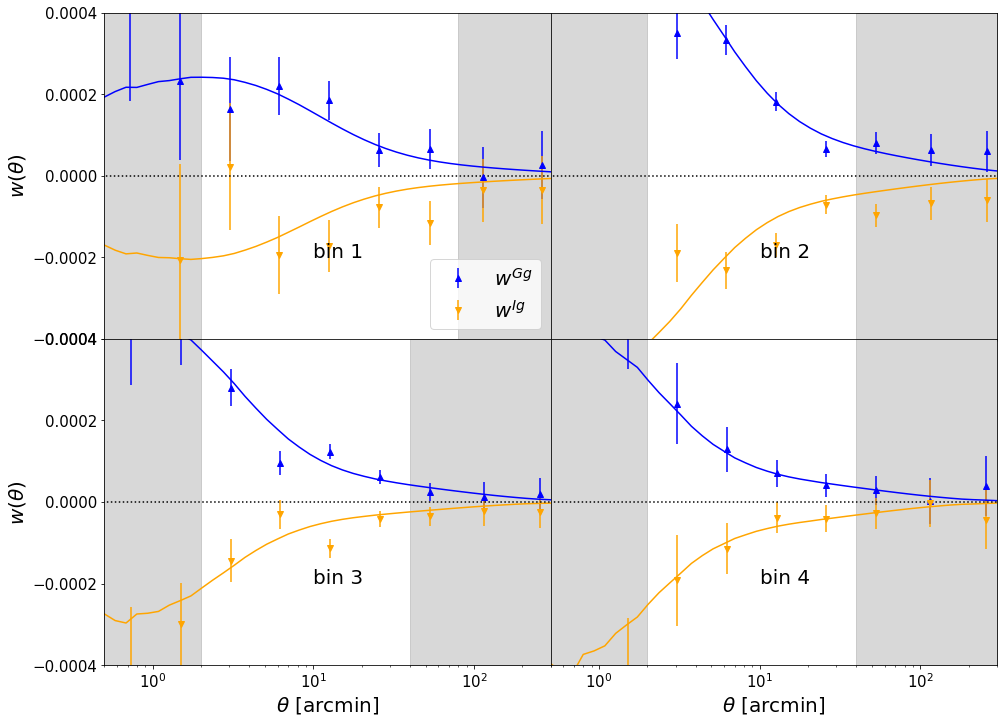}
		\caption{The lensing signal $w^{Gg}$ (blue up-triangles) and
			the IA signal $w^{Ig}$ (orange down-triangles) measured by the SC
			method. The grey shaded regions are the angular cuts where the effective $b_g(\theta)$ are not linear, see later in Fig.\,\ref{fig: effective bg red} for example. We also show the best-fit theoretical curves. In the fit,
			we fix cosmology, but varying the galaxy bias $b_g$ and the IA
			amplitude $A_{\rm IA}$ for the non-linear tidal alignment
			model. }
		\label{fig: lensing IA separation all}
	\end{figure}
	
	\begin{figure}\centering
		\hspace*{-0.1cm}
		%    \vspace*{-0.5cm}
		\includegraphics[width=1.0\columnwidth]{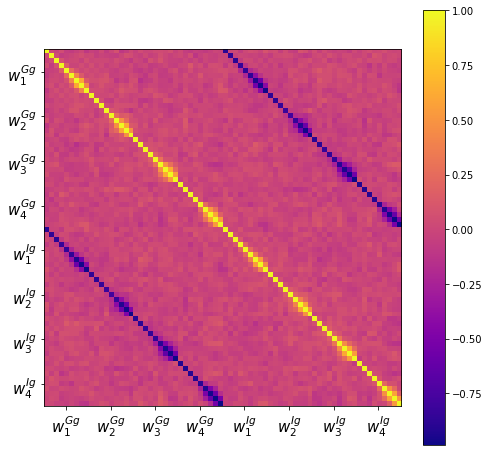}
		\caption{The normalized covariance matrix (The cross correlation
			coefficient) of the
			data vector ${\bf D}=(w^{Gg}(\theta), w^{Ig}(\theta))$. For each
			photo-z bin, there
			are 9 $\theta$-bins for $w^{Gg}$ and 9 for $w^{Ig}$, so the
			size for one z-bin is 18, and the overall size for the whole data vector is 72, leading to the
			$72\times72$ matrix above, corresponding to the
			combination of the 4 redshift bin shown in Fig.\,\ref{fig: lensing IA separation all}. The measured $w^{Gg}$ and $w^{Ig}$ show strong anti-correlation, which must be taken into account for quantifying the measurement significance and theoretical
			interpretation.}
		\label{fig: r_GgIg all}
	\end{figure}
	
	Now we compare with the theoretical prediction of the nonlinear tidal
	alignment model. Since the predicted $w^{Ig}\propto b_g A_{\rm IA}P_\delta$,
	we need to include the measurement $w^{Gg}\propto b_g P_\delta$, in order to break the
	$b_g$-$A_{\rm IA}$ degeneracy.  Since both $w^{Ig}$ and $w^{Gg}$ are derived from the same
	set of data, they are expected to have a strong negative
	correlation.  Fig.\,\ref{fig: r_GgIg all} confirms this expectation of
	strong anti-correlation. This figure shows the cross correlation
	coefficient (normalized covariance matrix),  $r_{ab}\equiv
	Cov(a,b)/\sqrt{Cov(a,a)Cov(b,b)}$. Here $a,b\in (w^{Gg}(\theta_1), w^{Gg}(\theta_2),\cdots,
	w^{Ig}(\theta_1), \cdots)$. Therefore we
	should fit for $w^{Gg}$ and $w^{Ig}$ simultaneously and take this
	anti-correlation into account.  We test that, if we ignore this
	strong anti-correlation and fit $w^{Gg}$ and $w^{Ig}$ separately, the
	bestfits do not well reproduce $w^{\gamma g}$ and $w^{\gamma g}|_S$ in Fig.\,\ref{fig: observables}. When doing the fitting, we only use the $34\times34$ matrix that correspond to the cuts in Fig.\,\ref{fig: lensing IA separation all}.
	
	We also notice the main correlation is between $w^{Gg}_i$ and $w^{Ig}_i$ in the same bin $i$. There is no significant correlation between different redshift bins. This is another proof that the impact from photo-z outlier to our lensing-IA separation is not significant.
	
	The theoretical fitting is carried out  with a fixed cosmology (Planck cosmology in Table
	\ref{table: fiducial cosmology}), and a fixed IA model (the nonlinear
	tidal alignment model). So there are only two free parameters in the
	fitting, namely the galaxy bias $b_g$ and the IA amplitude $A_{\rm
		IA}$. The two contain the leading order
	information of the measurements since $w^{Gg}\propto b_g$, and
	$w^{Ig}\propto b_gA_{\rm IA}$. Furthermore, a large fraction of
	cosmological dependence (in particular $\sigma_8$) can be absorbed into $b_g$ since both
	$w^{Gg}\propto b_gP_\delta$ and $w^{Gg}\propto (b_gP_\delta)\times
	A_{\rm IA}$.  Also for this reason, the constraint on $A_{\rm IA}$ is
	less cosmology-dependent than that on $b_g$. Since the major purpose
	of this work is to study IA,  the above simplification in model
	fitting meets our needs. With future data of significantly improved S/N, we will perform a global fitting with relaxed constraints
	of cosmology and IA models. 
	
	\begin{figure}\centering
		\hspace*{-0.3cm}
		\includegraphics[width=1.1\columnwidth]{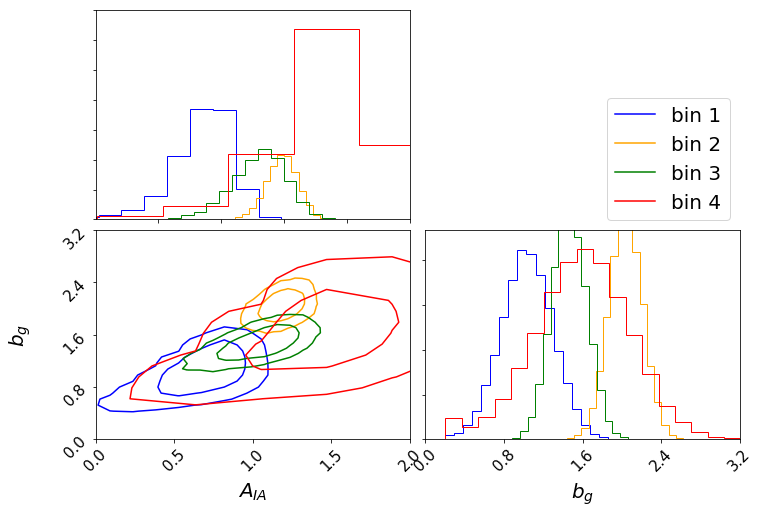}
		\caption{The MCMC fitting results (with 68\% and 95\% confidence contours) for the galaxy bias $b_g$ and IA
			amplitude $A_{\rm IA}$ of each photo-z bin. We find a clear redshift-dependent evolution on the IA amplitude $A_{\rm IA}$. The strong constraining power in bin 2 and 3 are due to their large numbers of galaxies, as shown in Fig.\,\ref{fig: n(z)}. The abnormal behavior of bin 2 is due to the large fraction of red galaxies and possible bias from photo-z, which will be discussed later in this work.}
		\label{fig: bg-IA fit all}
	\end{figure}
	
	\begin{figure}\centering
		\hspace*{-0.3cm}
		\includegraphics[width=1.1\columnwidth]{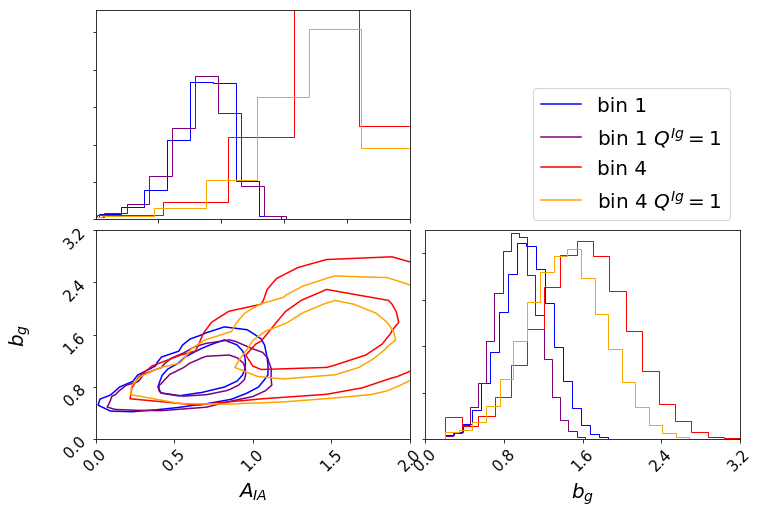}
		\caption{We show the comparison between using $Q^{Ig}(\theta)$ as in Eq.\,\ref{Eq Q^Ig(theta)} and assuming $Q^{Ig}=1$ as in previous work \citep{Yao2019}. The systematic error of assuming $Q^{Ig}=1$ is not significant for the current stage weak lensing surveys, however it could potentially matter for the stage IV surveys.}
		\label{fig: bg-IA fit all QIg}
	\end{figure}
	
	The MCMC fitting results on $b_g$ and $A_{\rm IA}$ are shown in
	Fig.\,\ref{fig: bg-IA fit all}, plotted with corner \citep{corner}. The best-fit values in this figure are
	used to plot the best-fit curves in Fig.\,\ref{fig: observables}
	and \ref{fig: lensing IA separation all}.  The best-fit
	curves agree with both the lensing signal and the IA signal reasonably well. This suggests that
	the LIS method works well, and support the non-linear tidal alignment
	IA model within the angular range of this work. In the future with better data and sufficient modeling of the small-scale, we can further investigate IA-physics in the non-linear regime.
	
	Fig.\,\ref{fig: bg-IA fit all} shows  a clear redshift-dependent on
	the IA amplitude $A_{\rm IA}$. Comparing with a redshift-independent fitting with the best-fit $A_{\rm IA}=1.05$, our measurements rule out the constant IA amplitude assumption at $\sim3\sigma$ (also see later in Fig.\,\ref{fig: IA(color,z)} with the $A_{\rm IA}(z)$ plot). When redshift increases, $A_{\rm IA}$
	becomes larger. The only exception is the redshift bin 2.  This is likely due to larger photo-z scatters and higher red galaxy fraction of the redshift bin 2. We will further discuss it in \S \ref{Section
	results red-blue}. We also investigated the impact of assuming $Q^{Ig}=1$ in Fig.\,\ref{fig: bg-IA fit all QIg}. We only show for bin 1 and 4 for readability, but we note that $A^{\rm IA}$ from this assumption is consistent with the ones with varying $Q^{Ig}(\theta)$.
	
	We caution that photo-z outlier can also lead to biased estimation in $A_{\rm IA}$. Even though this is beyond the scope of this paper, we try to quantify the quality of the photo-z being used in Appendix \ref{Apdx: photo-z}. More sanity checks will be discussed in the next section.
	
	The high S/N in Fig.\,\ref{fig: lensing IA separation
		all} motivates us to further investigate such following questions:
	\begin{enumerate}
		\item How do the IA signals depend on the galaxy color (red/blue
		galaxies) or other galaxy properties? 
		\item How does the IA amplitude evolves with redshift,  for red and blue galaxies?
		\item How good is the current non-linear tidal alignment model?
	\end{enumerate}

	\subsection{Separate IA measurements for red and blue galaxies} \label{Section results red-blue}
	
	\begin{figure}\centering
		\includegraphics[width=1.1\columnwidth]{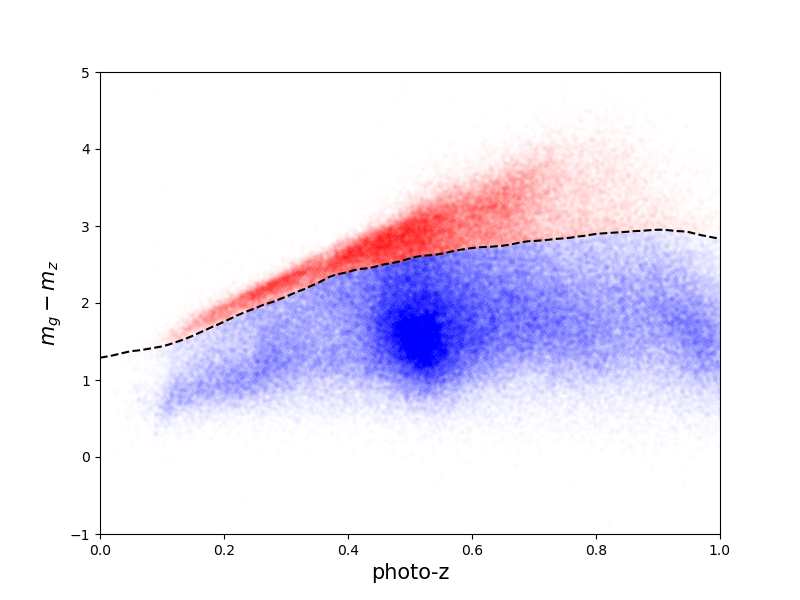}
		\caption{Red-blue galaxy classificatiopn through the
			color-redshift cut (black dashed curve) in the $m_g-m_z$
			v.s. $z^P$ space.  Table \ref{table: red-blue} shows the total
			number of red/blue galaxies}
		\label{fig: red-blue}
	\end{figure}
	
	\begin{table}\centering
		\caption{The number of red/blue galaxies, in the unit of millions
			(M). }
		\label{table: red-blue}
		%     \begin{ruledtabular}
		\begin{tabular}{ c c c c c c } 
			\hline
			& $0.1<z^P<0.9$ & z1 & z2 & z3 & z4  \\
			\hline
			Red+Blue & 23.4M  &  2.9M  &  6.1M  &  9.7M & 4.7M \\
			Red & 7.4M  &  0.8M  &  2.3M  &  3.2M & 1.1M \\
			Blue & 16.0M  &  2.0M  &  3.8M  &  6.5M & 3.6M \\
			Red fraction & 32\%  & 28\% & 38\% & 33\% & 23\% \\
			\hline
		\end{tabular}
		%     \end{ruledtabular}
	\end{table}
	
	The galaxy intrinsic alignment is expected to rely on galaxy type, and a major dependence is the galaxy color (red/blue galaxies).  Therefore we apply the SC method separately for red and blue galaxies. The classification is done through the estimated clustering
	effect in the color-redshift space, obtained with the kNN algorithm
	\citep{Zou2019}. The classification criteria are shown in
	Fig.\,\ref{fig: red-blue}, with the total number of red/blue galaxies shown
	in Table \ref{table: red-blue}. The overall red fraction is $32\%$.
	
	\begin{figure}\centering
		\hspace*{-0.5cm}
		\includegraphics[width=1.1\columnwidth]{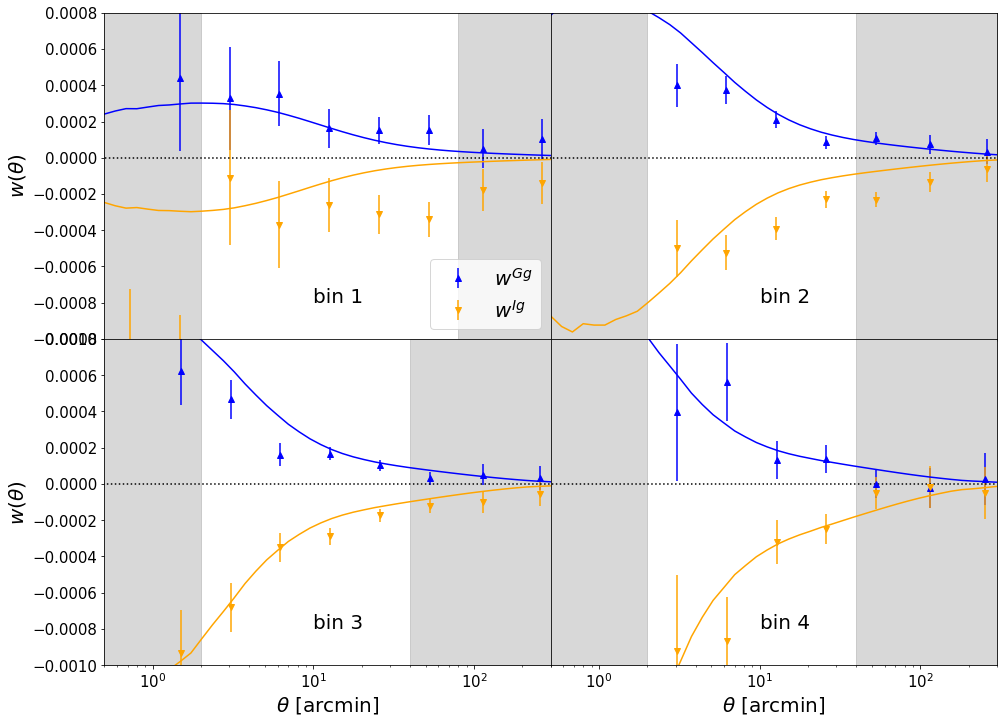}
		\caption{Similar to Fig. \ref{fig: lensing IA separation all}, but
			for red galaxies. The joint fit on the galaxy bias $b_g$ and the
			IA amplitude $A_{\rm IA}$ are shown in Fig.\,\ref{fig: bgIA
				red}, \ref{fig: IA(color,z)} \& Table \ref{table: AIA}.}
		\label{fig: GgIg red}
	\end{figure}
	
	\subsubsection{Red galaxies}
	Fig.\,\ref{fig: GgIg red} shows the separated lensing signal and IA
	signal for red galaxies, along with the best-fit theoretical curves. The detection of intrinsic alignment
	($w^{Ig}$) for red galaxies is significant at all four redshift bins and the
	corresponding S/N=$3.2$, $9.9$, $12.5$, $6.7$ respectively. Such S/N is comparable at low-z and significantly higher than the full sample at high-z, even with a much smaller sample (Table \ref{table: red-blue}). This means that blue galaxies included in the full sample contributes little to the IA signal, but induce significant noise and dilute the IA measurement S/N.  Generally, we achieved good fits for both the lensing part and the IA part. Overall the non-linear tidal alignment model is a good description to the IA of red galaxies. 
	
	\begin{figure}\centering
		\hspace*{-0.5cm}
		\includegraphics[width=1.1\columnwidth]{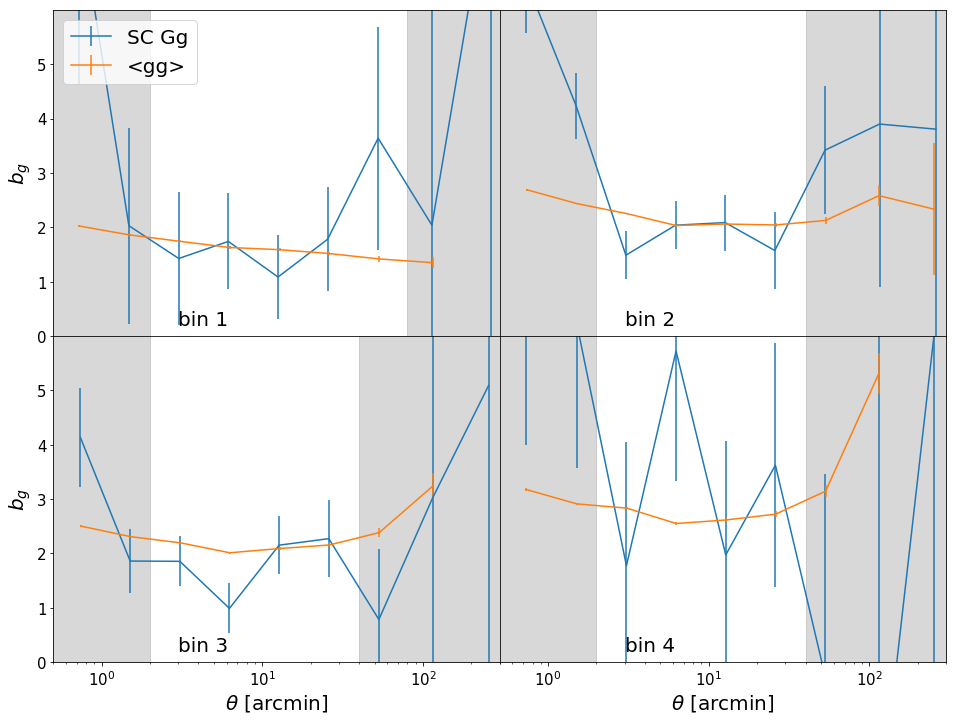}
		\caption{Comparison between effective galaxy bias $b_g$ from SC lensing signal ($b_g=w^{Gg}_{\rm SC}/w^{Gg}_{\rm theory,b=1}$, blue) and galaxy clustering ($b_g=\sqrt{w^{gg}_{\rm data}/w^{gg}_{\rm theory,b=1}}$, orange) for the red galaxies. The consistency between these two shows the accuracy of the lensing-IA separation.}
		\label{fig: effective bg red}
	\end{figure}
	
	We further present the effective galaxy bias obtained from the red galaxies for a sanity check in Fig.\,\ref{fig: effective bg red}. Since we have better S/N with red galaxies, it will be more important to show the consistent results from different methods. We get the effective galaxy bias from the SC-separated lensing signal, by calculating the ratio between the measurements from data and the theoretical predictions assuming $b_g=1$, namely $b_g=w^{Gg}_{\rm SC}/w^{Gg}_{\rm theory,b=1}$. Alternatively, it can be obtained from angular galaxy clustering of the same sample, following $b_g=\sqrt{w^{gg}_{\rm data}/w^{gg}_{\rm theory,b=1}}$. In Fig.\,\ref{fig: effective bg red} we showed these two methods give consistent results. This works as a further sanity-check in showing the results are robust against different systematics. For example:\\
	(1) the sharp non-linear galaxy bias is cut off at small-scales. \\
	(2) At large-scale when the effective $b_g$ is obviously non-linear, it could be the impact of the insufficient random catalog. Thus it is cut off.\\
	(3) Photo-z outlier should impact $w^{Gg}$ and $w^{gg}$ differently. While they are consistent, we know the impact from photo-z outlier is within reasonable range.
	
	\begin{figure}\centering
		\includegraphics[width=1.0\columnwidth]{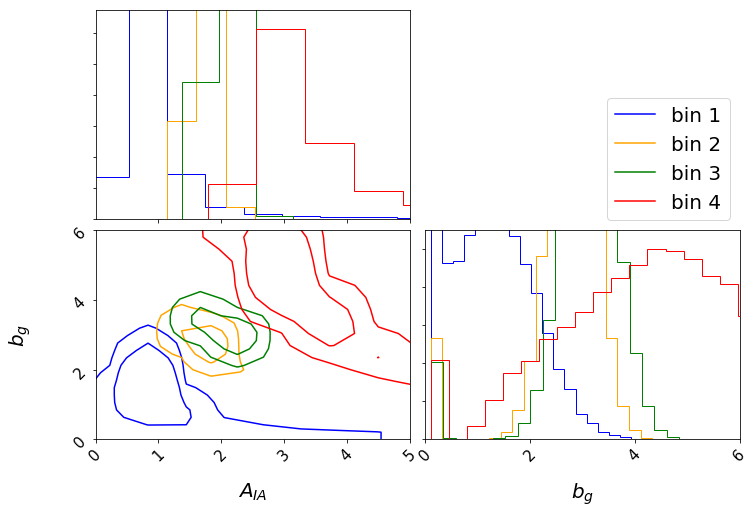}
		\caption{Similar to Fig. \ref{fig: bg-IA fit all}, but for red
			galaxies. We find a clear redshift-dependent evolution on the IA
			amplitude $A_{\rm IA}$. The overlap for the 2nd and 3rd redshift
			bins are likely due to significant overlap in their real redshift
			distribution.}
		\label{fig: bgIA red}
	\end{figure}
	
	Fig.\,\ref{fig: bgIA red} shows the
	constraints of $b_g-A_{\rm IA}$ for the red galaxies. We see a clear redshift evolution of
	$A_{\rm IA}$, namely $A_{\rm IA}$ increases with increasing $z$. Even for the 2nd and 3rd bins where the confidence contours are quite close, their $A_{\rm IA}$ differs at $\sim 2\sigma$ level, thanks to the small uncertainties from a large number of galaxies. Comparing with a redshift-independent fitting with the best-fit $A_{\rm IA}=1.87$, our measurements rule out the constant IA amplitude assumption at $\sim3.9\sigma$ (also see later in Fig.\,\ref{fig: IA(color,z)} with the $A_{\rm IA}(z)$ plot). For
	future cosmic shear or shear cross-correlation studies, it is then
	important to take this redshift dependence into account. This is also important in studies in galaxy formation, and it could be potentially related to \cite{Kurita2020}, where the halo IA (not the galaxy IA in our work) amplitude is also found to be z-dependent. The connection between halo IA and galaxy IA has also been discussed in \cite{Okumura2009}. More details about our IA results can be seen later in \ref{fig: IA(color,z)} \& Table \ref{table: AIA}
	
	Furthermore, recalling for the full (red+blue) sample, the second redshift bin has an unusually large $A_{\rm IA}$ (Fig. \ref{fig: bg-IA fit all}). The fact that the 2nd bin and 3rd bin have similar $A_{\rm IA}$ for the red galaxies may also be responsible in this situation.
	
	\begin{figure}\centering
		\hspace*{-0.5cm}
		\includegraphics[width=1.1\columnwidth]{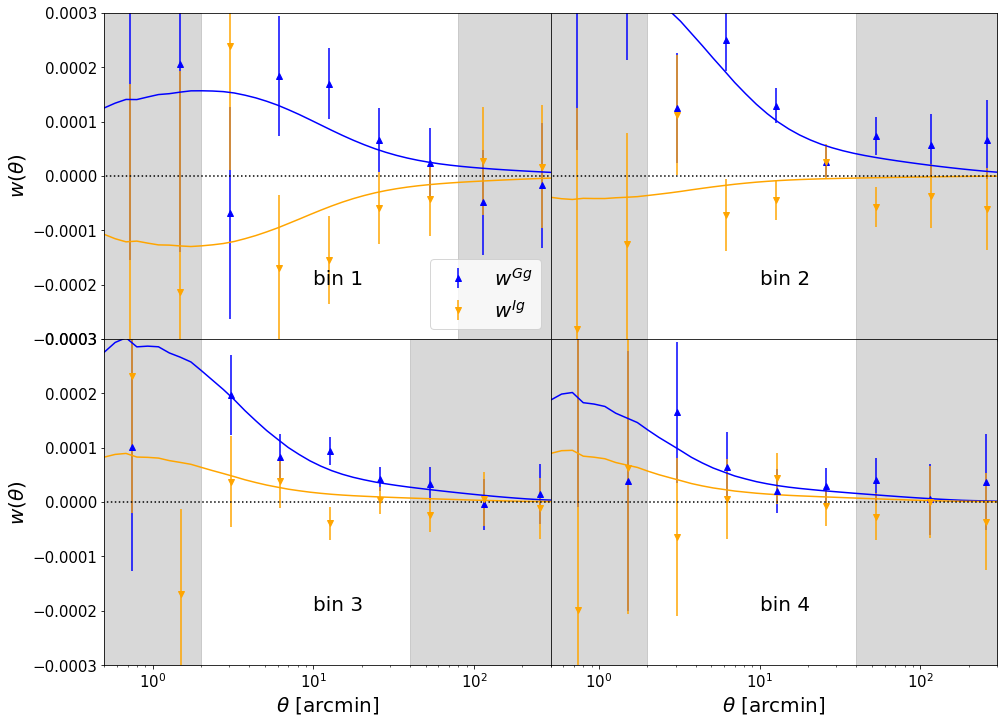}
		\caption{ Similar to Fig. \ref{fig: lensing IA separation all},
			but for blue galaxies. }
		\label{fig: GgIg blue}
	\end{figure}

	\subsubsection{Blue galaxies}
\begin{figure}\centering
		\includegraphics[width=1.0\columnwidth]{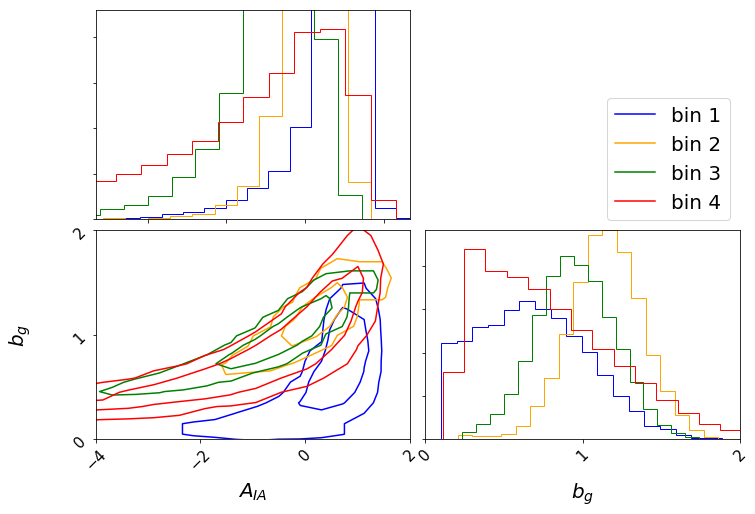}
		\caption{Similar to Fig. \ref{fig: bg-IA fit all}, but for blue
			galaxies. $A_{\rm IA}$ are consistent with 0.}
		\label{fig: bgIA blue}
	\end{figure}

	Fig.\,\ref{fig: GgIg blue} presents the separated lensing signal and
	IA signal, along with their best-fit theoretical curves, for blue galaxies. The $b_g$ and $A_{\rm IA}$ constraints are shown in
	Fig.\,\ref{fig: bgIA blue}, also later in Fig.\,\ref{fig: IA(color,z)} \& Table \ref{table: AIA}. Different from the red galaxies, we do not detect the IA signal in bins 2, 3, and 4. This generally agrees with our current understanding that the IA
	signals mainly exist in the red galaxies. However,  we do detect  IA
	signal for blue galaxies in the lowest redshift bin, although the signal is weak.  When fitted with the non-linear tidal alignment IA
	model, the detection significance is $\sim1\sigma$.  The current LIS
	method can not fully quantify the impact of photo-z outliers, plus blue galaxies normally have worse photo-z measurements comparing with red galaxies, therefore if this signal is real or not requires future exploration with better data.
	
	We note the results in Fig.\,\ref{fig: IA(color,z)} could still be affected by shear calibration bias and photo-z bias, with potential changes in the results quantified in Appendix \ref{Apdx: m and n(z) bias}. We expect in the future with a larger number of galaxies, better imaging and shear measurements, better photo-z, better modeling of other systematics (so that more information can be kept, instead of applying the scale-cuts), our SC method can further tell the physics for both red and blue galaxies. 
	
	% -----------------------------------
	\section{Summary and conclusions} \label{Section discussion}
	In this work, we apply the lensing-IA separation (LIS) pipeline of the
	self-calibration (SC) method to the DECaLS DR3 shear + photo-z
	catalog. This allows us to measure the galaxy intrinsic alignment signal, free of assumption on the IA model. Therefore the measurement not only reduces IA contaminations in weak lensing cosmology, but also provides valuable information on the physics of IA and galaxy formation.  Comparing
	to our previous work with the KiDS data \citep{Yao2019}, we have improved the technique and analysis over the following aspects:
	
	\begin{itemize}
		\item We improved the SC formalism with a scale-dependent $Q^{Gg}(\theta)$ rather than a constant, as in Eq.\,\eqref{Eq Q^Gg(theta)} and Fig.\,\ref{fig: Q}. This prevents a biased estimation of $w^{Gg}$ and $w^{Ig}$ at low-z that shifts power between large-scale and small-scale.
		
		\item We improved the SC formalism by introducing the IA-drop $Q^{Ig}\ne1$, due to non-symmetric redshift distribution, see Eq.\,\eqref{Eq Q^Ig(theta)} and Fig.\,\ref{fig: Q_Ig}. We showed in Fig.\,\ref{fig: bg-IA fit all QIg} that for the current stage the resulting $A_{\rm IA}$ is not biased even with the assumption $Q^{Ig}=1$. But it could matter for future surveys.
		
		\item We tested for different cosmology, as in Table \ref{table: fiducial cosmology}, the $Q$ parameter will be biased by $\sim10^{-5}$ to $\sim10^{-3}$, and the resulting $w^{Ig}$ will be biased by $\sim10^{-3}$ level. This demonstrated the bias from the fiducial cosmology that SC method need to assume is negligible. For the same reason, $Q^{Ig}$, by construct, is also insensitive to the assumed IA model. The bias from assumed IA model should be much smaller compared to Fig.\,\ref{fig: bg-IA fit all QIg}.
		
		\item We use jackknife resampling in each step of the calculation so that all the statistical uncertainties are included. We showed the statistical error on $Q$ is $\sim10^{-3}$ in Fig.\,\ref{fig: Q} and \ref{fig: Q_Ig}. This demonstrated our previous statement in \cite{Yao2017} that $Q$ won't introduce much statistical error. Addressing the systematic error from photo-z outlier, on the other hand, is beyond the scope of this paper as perfect knowledge on redshift is required.
		
		\item  We introduce the covariance between $w^{Gg}$ and $w^{Ig}$ in Fig.\,\ref{fig: r_GgIg all}, where the strong anti-correlation was not taken into consideration in previous work. This leads to more reliable fitting.
		
		\item We include the impact of galaxy bias $b_g$ in this work. It has been discussed to be one of the most important systematics in the SC method in \cite{Yao2017}. We performed a simultaneous fitting for the linear galaxy bias $b_g$ and IA amplitude $A_{\rm IA}$ to account for its effect, see in Fig.\,\ref{fig: bg-IA fit all}, \ref{fig: bgIA red} and \ref{fig: bgIA blue}.
		
		\item We apply additional scale cuts to prevent bias from different systematics, including non-linear galaxy bias, insufficient modeling of the matter power spectrum at small-scale, fake signal due to insufficient random catalog at large-scale, etc.
		
		\item We include multiple sanity checks in this work to validate our results, including checking the cross-shear (45-degree rotation) measurements are consistent with 0, comparing the resulting effective galaxy bias between the separated $w^{Gg}$ and galaxy clustering $w^{gg}$, no significant correlation between different z-bins in the covariance matrix, comparing $A_{\rm IA}$ with other analysis, etc.
	\end{itemize}
	
	\begin{table}\centering
		\caption{The best-fit $A_{\rm IA}$ and the $1\sigma$ error.}
		\label{table: AIA}
		%     \begin{ruledtabular}
		\begin{tabular}{ c c c c c } 
			\hline
			$A_{\rm IA}$ & z1 & z2 & z3 & z4  \\
			\hline
			Red+Blue & $0.70^{+0.15}_{-0.20}$  &  $1.19^{+0.10}_{-0.10}$  &  $1.05^{+0.15}_{-0.19}$  &  $1.47^{+0.25}_{-0.36}$  \\
			Red & $0.82^{+0.41}_{-0.26}$  &  $1.69^{+0.19}_{-0.17}$  &  $2.00^{+0.19}_{-0.16}$  &  $3.06^{+1.00}_{-0.46}$  \\
			Blue & $0.69^{+0.28}_{-0.59}$ &  $0.18^{+0.37}_{-0.52}$  & $-0.49^{+0.64}_{-1.03} $  &  $-0.75^{+1.32}_{-3.08}$  \\
			\hline
		\end{tabular}
		%     \end{ruledtabular}
	\end{table}
	
	\begin{figure}
		\centering
		\includegraphics[width=1.0\columnwidth]{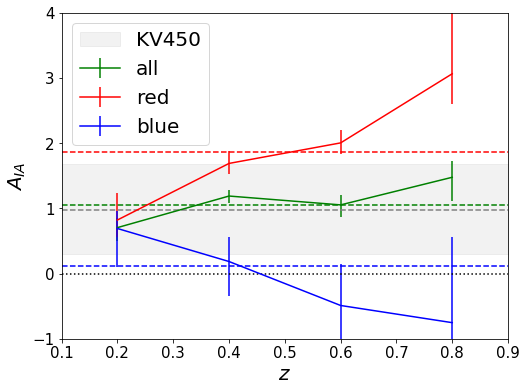}
		\caption{The color- and redshift-dependence of the best-fit $A_{\rm IA}$. Dashed lines are the best-fit with the constant $A_{\rm IA}$ assumption.}
		\label{fig: IA(color,z)}
	\end{figure}
	
	With the above improvements, we obtain reliable measurements on the separated lensing signal $w^{Gg}$ and IA signal $w^{Ig}$. Our findings can be summarized in Table \ref{table: AIA} and \ref{table: chi2} and visualized in Fig.\,\ref{fig: IA(color,z)}, with the following aspects:
	\begin{itemize}
		\item The separation and measurement of lensing and IA are more robust and statistically significant. A crucial diagnostic is the differences in the two direct observables $w^{\gamma g}$ and $w^{\gamma g}|_S$. The measured difference is improved to $\sim16\sigma$ for a single redshift bin (bin 2) and $\sim21\sigma$ (comparing with $\sim16\sigma$ in our previous work \cite{Yao2019}) for the full galaxy sample.  For this reason, the total detection significance of the IA signal reaches $\sim14\sigma$. The overall IA amplitude of our DECaLS DR3 sample is consistent with the KV450 \citep{Hildebrandt2018} results, but with stronger constraint, see in Fig.\,\ref{fig: IA(color,z)}. It is also consistent with the common understanding that $A_{\rm IA}\sim1$.
		
		\item We detect the IA dependence on galaxy color. For red galaxies, we detect IA in all photo-z bins at $0.1<z^P<0.9$. The detected IA signal shows reasonable agreement with the nonlinear tidal alignment model. The red-blue separation increases the S/N of IA detection in red galaxies to $\sim17.6\sigma$.
		
		\item We find for blue galaxies, the IA signal is generally consistent with 0, except for the weak and tentative ($\sim 1\sigma$) detection in the lowest redshift bin at $z^P<0.3$.
		
		\item Our results rule out the assumption of constant IA amplitude at $\sim3.9\sigma$ for the red sample, and at $\sim3\sigma$ for the full sample. Especially for red galaxies, the IA amplitude $A_{\rm IA}$ increases with redshift. From Fig.\,\ref{fig: IA(color,z)} we can also see a (not clear) evolution pattern for the blue galaxies, nonetheless, the full sample also seems to have a $A_{\rm IA}(z)$ evolution pattern, which agrees with our previous finding for KV450 \citep{Yao2019} that IA is stronger at high-z. Tests on how the calibrations of multiplicative bias and redshift distribution can affect the $A_{\rm IA}(z)$ relation are shown in Appendix \ref{Apdx: m and n(z) bias}. More test on the z-dependencies for the full sample and the blue sample can be done with a larger galaxy number and better photo-z in the future. We note similar z-evolution results has been found in a recent study with hydrodynamic simulations \citep{Samuroff2020}.
		
		\item Our separated IA signals do not rely on strong assumptions about IA physics. The MCMC fitting for $b_g$ and $A_{\rm IA}$ assumed the non-linear tidal alignment model, also known as the non-linear linear alignment (NLA) model, see in Eq.\,\eqref{Eq IA 3D}. But it can also be used to investigate other alternatives, for example \cite{Blazek2017,Fortuna2020}. Here we present the fitting $\chi^2$ in Table \ref{table: chi2}. We notice that for the red galaxies, in bin 2 and bin 3 where the IA detection is most significant, the $\chi^2/d.o.f.$ is not ideal. This suggests possible systematics and/or potential deviation from the assumed NLA model. However, the relatively large $\chi^2$ could also come from photo-z outlier (see Appendix \ref{Apdx: photo-z}) that we are unable to fully address in this work. We leave this point for future studies.
	\end{itemize}
	
	\begin{table}\centering
		\caption{Goodness of fit ($\chi^2$) to the measured $w^{Gg}$ and
			$w^{Ig}$ by the nonlinear tidal alignment model. The large
			$\chi^2$ mainly arises from $\lesssim 5$ Mpc scale
			(Fig.\,\ref{fig: lensing IA separation all}, \ref{fig: GgIg red}
			and \ref{fig: GgIg blue}). They suggest improvement in the
			theoretical modelling by taking complexities such as baryonic
			physics, non-linear galaxy bias and beyond tidal alignment  IA
			models into account. }\label{table: chi2}
		%     \begin{ruledtabular}
		\begin{tabular}{ c c c c c } 
			\hline
			$\chi^2/d.o.f.$ & z1 & z2 & z3 & z4  \\
			\hline
			Red+Blue & 22.4/8 & 32.2/6 & 20.2/6 &
			3.6/6 \\
			Red & 27.5/8 & 72.0/6 & 68.0/6 &
			3.6/6 \\
			Blue & 7.4/8 & 7.0/6 & 4.6/6 & 3.5/6 \\
			\hline
		\end{tabular}
		%     \end{ruledtabular}
	\end{table}
	
	With better data such as DECaLS DR8, future data release from KiDS/HSC/DES/LSST/etc, and possible improved photo-z estimation and shear measurements (which are beyond the scope of this paper, see discussions in Appendix \ref{Apdx: m and n(z) bias}), we plan to robustly measure the IA amplitude, and its dependence on the physical scale, redshift and galaxy properties such as color and flux. We may also be able to
	reveal more detailed information, such as the observed negative
	$b_g$-$A_{\rm IA}$ correlation in red galaxies, and the possibly
	positive correlation in blue galaxies (Fig. \ref{fig: bgIA red} \& \ref{fig: bgIA blue}).  This information will be useful to understand galaxy formation. Furthermore, the same analysis also provides the measurement of $w^{Gg}$,  namely the lensing-galaxy cross-correlation free of IA
	contaminations. This data contains useful information to constrain cosmology, as discussed in the previous work \citep{Yao2019}. This method could also potentially be affected by modified gravity, as the separated lensing signal relies on the gravitational potential $\nabla^2(\phi-\psi)$, while the IA signal relies on $\nabla^2\phi$ \citep{Zhang2007}. We will present more cosmological studies in separate future works.

	% -------------------------------------

	\section{acknowledgements}
	The authors thank the referee for many useful comments, which highly improved the quality of this paper. The authors thank Hu Zou, Haojie Xu, Jiaxin Wang, Minji Oh, Zhaozhou Li for useful discussions. JY and PZ acknowledge the support of the National Science Foundation of China (11621303, 11433001). HYS acknowledges the support from NSFC of China under grant 11973070, the Shanghai Committee of Science and Technology grant No.19ZR1466600 and Key Research Program of Frontier Sciences, CAS, Grant No. ZDBS-LY-7013. The computations in this paper were run on the $\pi$ 2.0 cluster supported by the Center for High Performance Computing at Shanghai Jiao Tong University.
	
	The codes JY produced for this paper were written in Python. JY thanks all its developers and especially the people behind the following packages: SCIPY \citep{scipy}, NUMPY \citep{numpy}, ASTROPY \citep{astropy} and MATPLOTLIB \citep{matplotlib}.

	% ------------------------------------
	
	\bibliography{references}
	\bibliographystyle{aasjournal} %apsrev4-1
	
	\appendix
	
	\section{Validating the photo-z quality} \label{Apdx: photo-z}
	We emphasize that the photo-z techniques are beyond the scope of this paper. Nonetheless, here we present the validation of the photo-z samples being used in this work, in addition to the correlation functions. We combine galaxies from UDS HSC + SPLASH \citep{Mehta2018}, ECDFS \citep{Cardamone2010}, CFHTLS Deep + WIRDS \citep{Bielby2010}, and COSMOS \citep{Laigle2016}, to get a large reliable photo-z catalog. The overall redshift distribution is quite similar to the $n(z)$ determined from COSMOS only, and was already presented in \cite{Phriksee2019}. By matching the above ``good photo-z catalog'' with our catalog of DR3 shear and kNN photo-z, we have a resulting sample with 46961 galaxies.
	
\begin{figure}\centering
		\includegraphics[width=0.5\columnwidth]{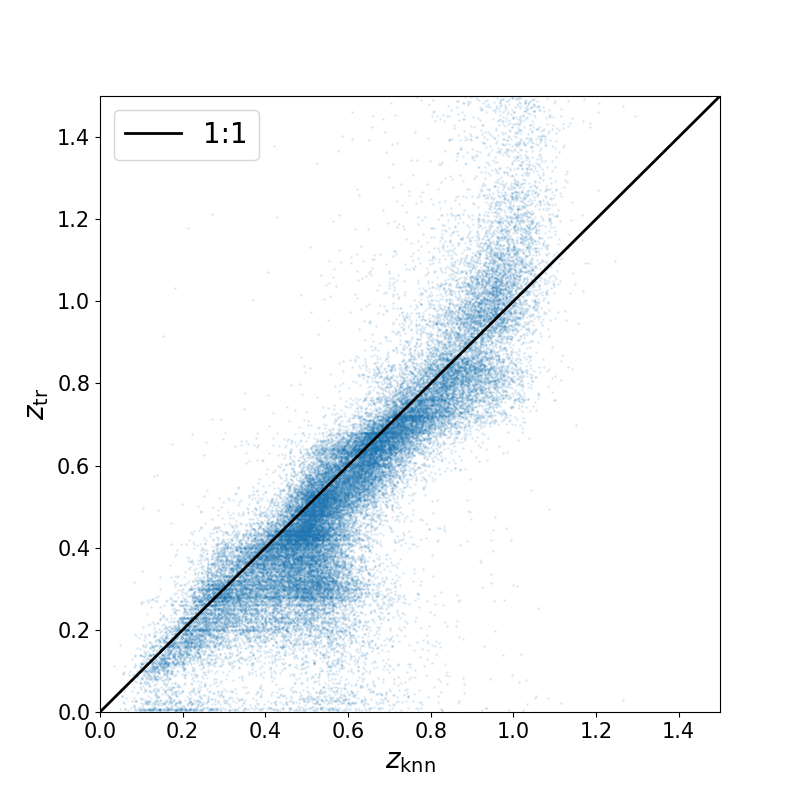}
		\caption{In this figure we present the comparison between the kNN photo-z ($z_{\rm knn}$ shown on the x-axis) samples being used in this work, and the selected good redshift ($z_{\rm tr}$ shown on the y-axis) samples. There are clearly two outlier regions, at $z_{\rm knn}\sim0.5$ (corresponding to mainly bin 2 and 3 of this work) and $z_{\rm knn}\sim 1$ (which is cut off in this work). We calculated the photo-z outlier rate $f_{\Delta z>0.15}$, defined as the fraction with $|z_{\rm knn}-z_{\rm tr}|>0.15$, which are [0.09, 0.19, 0.26, 0.15] for the 4 z-bins being used. The corresponding systematic shift $<z_{\rm knn}-z_{\rm tr}>$ are [0.02, 0.06, 0.08, 0.01].}
		\label{fig: photo-z 1:1}
	\end{figure}

	We refer to the ``good photo-z catalog'' as ``true-z'' in the following tests. In Fig.\,\ref{fig: photo-z 1:1} we present the direct comparison between the kNN photo-z $z_{\rm knn}$ \citep{Zou2019} in this work and the ``true-z'' $z_{\rm tr}$ described above. There are two regions that deviate from the 1:1 line significantly. The one we don't need to care about is the outlier region with $z_{\rm knn}\sim1$, since it has been cut off with our binning selection $0.1<z_{\rm knn}<0.9$. The outlier region we need to care about is $z_{\rm knn}\sim0.5$. The main photo-z outlier will be affecting bin 2 and 3, causing some disorder in the estimated photo-z and biasing the resulting $w^{Gg}$ and $w^{Ig}$ measurements. We think the high outlier rates and systematic shifts in bins 2 and 3 correspond to the high $\chi^2$ values shown in Table \ref{table: chi2}. On the other hand, the relatively reliable photo-z in bin 1 and 4 justified our result of IA redshift evolution.
	
\begin{figure}\centering
		\includegraphics[width=0.5\columnwidth]{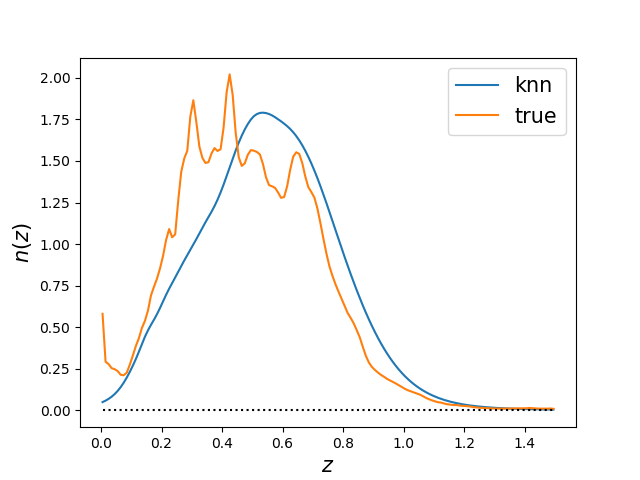}
		\caption{In this figure we show the redshift distribution $n(z)$ for the kNN photo-z (labelled as ``knn'') and the good redshift samples (labelled as ``true''). In general, the two curves have similar scatter and very close amplitudes, demonstrating that the Gaussian scatter given by the kNN photo-z is applicable. However, significant disagreement is showing at $z\sim0.4$, which corresponds to some redshift outlier problem in bin 2 of this work. We also notice that the kNN photo-z seems to be systematically higher.}
		\label{fig: photo-z n(z)}
	\end{figure}

	We further present the redshift distribution of this work and the reference ``good photo-z sample'' in Fig.\,\ref{fig: photo-z n(z)}. The $n(z)$ used in this work is shown as the ``knn'' distribution, which has very similar amplitude and scatter comparing with the reference ``true'' $n(z)$. This demonstrates that the given Gaussian redshift scatter from \cite{Zou2019} is generally reasonable. On the other hand, we do observe a significant difference at $z\sim0.4$, resulting from the significant outlier problem shown previously in Fig.\,\ref{fig: photo-z 1:1}. This also agrees with the arguments in \cite{Zou2019} that the main redshift-color degeneracy will happen in this redshift range, leading to some misclassification of the photo-z.
	
	Generally, the photo-z quality in this work is suitable for the study of self-calibration. The kNN photo-z \citep{Zou2019} gives reliable best-fit photo-z and Gaussian scatter to present the underlying $n(z)$. However, we found that due to the redshift-color degeneracy discussed in \cite{Zou2019}, there are some significant redshift outliers in our bin 2 and 3, which can lead to some bias in our $w^{Gg}$ and $w^{Ig}$. This bias is smaller for red galaxies as their photo-z is generally better. There could also be biases due to training sample selection, for example \cite{Hartley2020}, but they are beyond the scope of this paper. 
	
	\section{Calculating the lensing-drop and IA-drop $Q$} \label{Apdx: eta}
	
	The lensing-drop $Q^{Gg}$ and the IA-drop $Q^{Ig}$ play crucial roles in lensing-IA separation (Eq.\,\eqref{Eq Gg correlation} \& \eqref{Eq Ig correlation}), where \{$w^{Gg}$, $w^{Ig}$\} comes from Hankel transformation as in Eq.\,\eqref{Eq Hankel}. Therefore to get the $Q$s, we need to calculate the power spectra for \{$C^{Gg}$, $C^{Gg}|_S$, $C^{Ig}$, $C^{Ig}|_S$\}, with the given photo-z information of the survey.
	
	Theoretically, $C^{Gg}_{ii}$ is given by Eq. \eqref{Eq Gg}, and
	$C^{Gg}_{ii}|_S$ is given by 
	\begin{equation}
	C^{Gg}_{ii}|_S(\ell)=\int_0^\infty\frac{W_i(\chi)n_i(\chi)}{\chi^2}
	b_g P_\delta\left(k=\frac{\ell}{\chi};\chi\right) \eta^{Gg}_i(z) d\chi
	. 
	\label{Eq GgS}
	\end{equation}
	The extra factor $\eta^{Gg}_i(z)$ arises from the fact that  $C^{Gg}|_S$
	only contains pairs with $z^P_\gamma<z^P_g$ \citep{SC2008}. 
	\ba
	\eta^{Gg}_i(z)&=&\eta^{Gg}_i(z_L=z_g=z)\ ,  \\
	\eta^{Gg}_i(z_L,z_g) &=&\frac 
	{2\int dz^P_{G}\int dz^P_g\int_{0}^{\infty}dz_G W_L(z_L,z_G)p(z_G|z^P_G)p(z_g|z^P_g)S(z^P_G,z^P_g)n^P_i(z^P_G)n^P_i(z^P_g)}
	{\int dz^P_{G}\int dz^P_g\int_{0}^{\infty}dz_G
		W_L(z_L,z_G)p(z_G|z^P_G)p(z_g|z^P_g)n^P_i(z^P_G)n^P_i(z^P_g)}\ .
	\nonumber\label{Eq eta}
	\ea
	Here $z_L$, $z_g$ and $z_G$ denote the lens redshift, the galaxy
	redshift, and the lensing source redshift, respectively. The quantities
	with superscript ``P'' denote photometric redshifts $z^P$ and the ones
	without it are the true redshifts $z$. The  integral $\int dz_G^P$
	and $\int dz_g^P$ are both over $[z^P_{i,\rm min}, z^P_{i,\rm max}]$,
	namely the photo-z range of the $i^{\rm th}$ tomographic bin. The lensing
	kernel $W_L$ for a flat universe is given by 
	\begin{equation}
	W_L(z_L,z_S)=\begin{cases}
	\frac{3}{2}\Omega_m\frac{H_0^2}{c^2}(1+z_L)\chi_L(1-\frac{\chi_L}{\chi_S}) &\text{for $z_L<z_S$}\\
	0 &\text{otherwise}
	\end{cases};
	\end{equation}
	$p(z|z^P)$ is the redshift probability distribution function (PDF). In
	reality each galaxy has its own PDF. To speed up the calculation, we
	approximate it as a Gaussian function identical for all galaxies
	with the same $z^P$, as we adopted in the previous work
	\citep{Yao2017}. 
	$S(z^P_G,z^P_g)$ is the selection function for the ``$|_S$'' symbol, 
	\begin{equation} \label{Eq selection}
	S(z^P_G,z^P_g)=\begin{cases}
	1 &\text{for $z^P_G<z^P_g$}\\
	0 &\text{otherwise}\ .
	\end{cases};
	\end{equation}
	$n^P_i(z^P)$ gives the photo-z distribution function in the
	$i^{\text{th}}$ tomographic bin.  The  calculation of $\eta(z)$ can be
	extremely massive, since different galaxies (even with the same $z^P$)
	in general have different  photo-z PDF. For fast calculation, we
	follow our previous work \citep{Yao2017} and 
	assume a uniform Gaussian PDF for all galaxies in the given photo-z
	bin,
	\begin{equation}
	p(z|z^P)=\frac{1}{\sqrt{2\pi}\sigma_z(1+z)}{\rm
		exp}\bigg\{-\frac{(z-z^P-\Delta_z^i)^2}{2[\sigma_z(1+z)]^2}\bigg\}\
	. \label{Eq Gaussian PDF}
	\end{equation}
	$\sigma_z$ in the above equation is  the averaged photo-z scatter of
	all galaxies  in the given photo-z bin. This assumption is valid because the redshift Gaussian scatter is tested in the machine learning method \citep{Zou2019} and is also checked in Fig.\,\ref{fig: photo-z n(z)} as they have similar height and scatter compared to the ``true-z'', despite of the outlier problem. 
	
	The factor $2$ in Eq.\,\eqref{Eq eta} arises from
	an integral equality theoretically predicted in \citet{SC2008},
	\ba
	\label{Eq 2}
	\frac{\int_{z^P_{ i, \rm min}}^{z^P_{ i, \rm max}}dz^P_{G}\int_{z^P_{ i, \rm min}}^{z^P_{ i, \rm max}}dz^P_g 
		n^P_i(z^P_G)n^P_i(z^P_g)}{\int_{z^P_{ i, \rm min}}^{z^P_{ i, \rm max}}dz^P_{G}\int_{z^P_{ i, \rm min}}^{z^P_{ i, \rm max}}dz^P_g 
		n^P_i(z^P_G)n^P_i(z^P_g)S(z^P_G,z^P_g)}=2\ .
	\ea
	This has also been tested numerically.
	
	The $Q^{Ig}$ introduce in this paper share similar definition as above. $C^{Ig}_{ii}$ is defined in Eq.\,\eqref{Eq Ig}, while $C^{Ig}_{ii}|_S$ is defined as
	\begin{equation}
	C^{Ig}_{ii}|_S(\ell)=\int_0^\infty\frac{n_i(\chi)n_i(\chi)}{\chi^2}b_gP_{\delta,\gamma^I}\left(k=\frac{\ell}{\chi};\chi\right)\eta^{Ig}_i(z) d\chi, \label{Eq IgS}
	\end{equation}
	in which $\eta^{Ig}$ is given by
	\begin{equation}
	\eta^{Ig}_i(z_L,z_g) =\frac 
	{2\int dz^P_{G}\int dz^P_g\int_{0}^{\infty}dz_G p(z_G|z^P_G)p(z_g|z^P_g)S(z^P_G,z^P_g)n^P_i(z^P_G)n^P_i(z^P_g)}
	{\int dz^P_{G}\int dz^P_g\int_{0}^{\infty}dz_G
		p(z_G|z^P_G)p(z_g|z^P_g)n^P_i(z^P_G)n^P_i(z^P_g)}\,
	\nonumber\label{Eq eta Ig}
	\end{equation}
	simply without the lensing kernel $W_L(z_L,z_S)$ comparing to $\eta^{Gg}$, as the I-g correlation differs from the G-g correlation.
	
	The calculation of \{$Q^{Gg}(\theta)$, $Q^{Ig}(\theta)$\} requires the photo-z distribution
	$n^P_i(z^P)$, the true redshift distribution $n_i(z)$, and  cosmology
	(e.g. through $P_\delta$ and $W_L(z_L,z_S)$).  However, its
	cosmological dependence is weak, since the cosmology dependent terms enter the same way in both $C^{Gg}$ and $C^{Gg}|_S$ and therefore largely cancel each other in the ratio ($Q$). We tested for different cosmology in Table \ref{table: fiducial cosmology}, the difference is at $\sim 10^{-3}$ to $\sim 10^{-5}$ level for Q. With the development in this paper, we also show the relation of power spectra based $Q(\ell)$ and correlation function based $Q(\theta)$ in Fig.\,\ref{fig: Q} and \ref{fig: Q_Ig}.

	\section{Potential biases from shear measurements and redshift distribution} \label{Apdx: m and n(z) bias}
	\begin{figure}\centering
		\includegraphics[width=0.5\columnwidth]{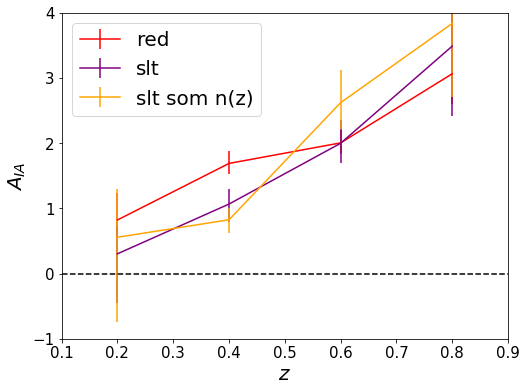}
		\caption{Test the $A_{\rm IA}(z)$ relation with the default cut (with ``SIMP'' galaxies and $1+m>0.5$) in red, with the selection for m-bias (without ``SIMP'' and without $1+m>0.5$) in purple, and impact from different $n(z)$ estimation in orange.}
		\label{fig: m and n(z) bias}
	\end{figure}

	To further validate our results, we investigate the impact of (1) bias from shear calibration, and (2) bias from redshift distribution $n(z)$ calibration. We choose to use the red galaxies as an example, since its IA redshift-dependency $A_{\rm IA}(z)$ in Fig.\,\ref{fig: IA(color,z)} is the most delicate result of this work. We note accurate calibration in either shear \citep{Pujol2020,Huff2017,Sheldon2017} or redshift \citep{Hildebrandt2020,Hildebrandt2016} are beyond the scope of this paper.
	
	To have a better assessment on the impact of biased multiplicative bias $m$, we choose to cut off the ``SIMP'' type galaxies as in \cite{Phriksee2019}. The shape measurements of ``SIMP'' subsample are quite noisy, so they contribute less in our results. Meanwhile by removing this type of galaxies, the remaining sample is dominated by the ``EXP'' type ($>80\%$), whose multiplicative bias is accurately estimated as shown in Table A1 of \cite{Phriksee2019}.
	We no longer use the default cut of $1+m>0.5$ as it could also introduce some selection bias. The associated results are shown in Fig.\,\ref{fig: m and n(z) bias} with label ``slt''. Although the values of the IA amplitude $A_{\rm IA}$ changed slightly in each z-bin, we note by selecting different types of galaxies, the selected galaxies should have different IA amplitude. We emphasize that the IA redshift-evolution result of the red galaxies remains the same: it rules out the constant IA amplitude assumption at $\sim4\sigma$ (with slightly larger errorbars but reduced amplitude in the low-z bins, comparing with the default red sample).
	
	We further address the impact of biased $n(z)$ estimation in the theoretical part. We match the ``good photo-z catalog'' described above (0.69M galaxies) with the ``slt'' red galaxy sample, resulting in 6.5k matches. Instead of using the $n(z)$ given by the kNN photo-z, we choose to use the distribution from the matched ``good photo-z catalog'' in the theoretical calculation of Eq.\,\eqref{Eq Gg} and \eqref{Eq Ig}. In this way, the IA signal gives the orange ``slt som n(z)''-labeled results in Fig.\,\ref{fig: m and n(z) bias}. Still, it shows the impact from the biased $n_i(z)$ is not significant, and the redshift-evolution result of the red galaxies remains the same. We note that getting $n(z)$ with another catalog will also add an extra selection bias on the IA amplitude.
	
	Even though the above two tests on shear calibration and redshift distribution calibration are not required to produce the same results as our default red galaxy sample, they still agree at some level. Therefore we conclude that the $A_{\rm IA}(z)$ relation found for the red galaxies is robust against the described calibration biases.
	
	\section{Covariance matrix for the observables}
	\label{Apdx: cov observable}
	\begin{figure}\centering
		\includegraphics[width=0.5\columnwidth]{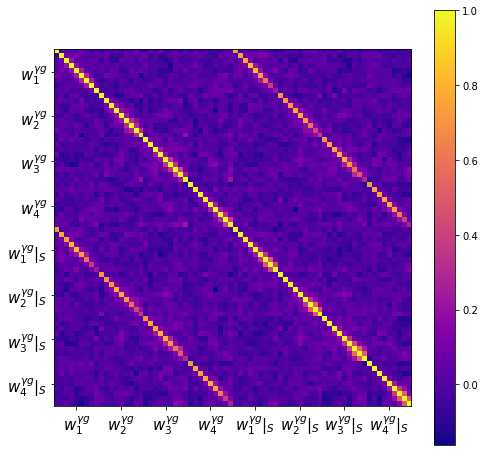}
		\caption{The normalized covariance matrix (the correlation
			coefficient $r_{ab}=Cov(a,b)/\sqrt{Cov(a,a)Cov(b,b)}$) for the
			LIS observable data vector \{$w^{\gamma g}(\theta)$, $w^{\gamma
				g}|_S(\theta)$\}. There are 9 $\theta$-bins for $w^{\gamma
				g}$ and 9 for $w^{\gamma g}|_S$, so the overall size for the
			data vector is 18 for each z-bin, leading to the $72\times72$ matrix for the full sample above. There are strong positive
			correlation between $w^{\gamma g}$ and $w^{\gamma g}|_S$,
			important for the data analysis. }
		\label{fig: r_obs all}
	\end{figure}
	
	We show the normalized covariance matrix of \{$w^{\gamma g}(\theta)$,
	$w^{\gamma g}|_S(\theta)$\} in Fig.\,\ref{fig: r_obs all}. It is
	obvious that the two observables have a strong positive correlation,
	simply due to the fact that the data producing $w^{\gamma
		g}|_S(\theta)$ is completely included in $w^{\gamma
		g}(\theta)$. This positive correlation is converted into a negative
	correlation in the separated $w^{Gg}$ and $w^{Ig}$ (Fig. \ref{fig: r_GgIg all}), through our lensing-IA separation method in Eq.\,\eqref{Eq Gg correlation} and \eqref{Eq Ig correlation}. The only difference is the covariance of \{$w^{Gg}, w^{Ig}$\} contains the statistical uncertainties from \{$Q^{Gg}$, $Q^{Ig}$\}, which we tested to be at $\sim10^{-3}$ level. So generally Fig.\,\ref{fig: r_GgIg all} and \ref{fig: r_obs all} carries equivalent information.
	
	\label{lastpage}
\end{document}